\documentclass[12pt,letterpaper]{article}
\pdfoutput=1

\usepackage{amsmath,amssymb,calc, amsthm}
\usepackage{graphicx}
\usepackage[nosort]{cite}
\usepackage{epsfig,psfrag}

\makeatletter
\renewcommand\section{\@startsection {section}{1}{\z@}%
                                   {-3.5ex \@plus -1ex \@minus -.2ex}
                                   {2.3ex \@plus.2ex}%
                                   {\normalfont\large\bfseries}}
\renewcommand\subsection{\@startsection{subsection}{2}{\z@}%
                                     {-3.25ex\@plus -1ex \@minus -.2ex}%
                                     {1.5ex \@plus .2ex}%
                                     {\normalfont\bfseries}}
\makeatother

\theoremstyle{plain}

\theoremstyle{definition}

\let\non\nonumber

\def\one{^{(1)}}

\newcommand{\bea}{\begin{eqnarray}}
\newcommand{\eea}{\end{eqnarray}}
\newcommand{\be}{\begin{equation}}
\newcommand{\ee}{\end{equation}}
\newcommand{\bma}{\begin{pmatrix}}
\newcommand{\ema}{\end{pmatrix}}
\newcommand{\cc}{{\rm c.c.}}

\newcommand{\hlf}{\frac{1}{2}}

\newcommand{\Z}{{\mathbb Z}}

\newcommand{\com}[2]{{ \left[ #1, #2 \right] }}
\newcommand{\acom}[2]{{ \left\{ #1, #2 \right\} }}

\def\com#1#2{{ \left[ #1, #2 \right] }}
\def\acom#1#2{{ \left\{ #1, #2 \right\} }}

\newcommand{\C}[1]{$(\ref{#1})$}

\def\IZ{\relax\ifmmode\mathchoice
{\hbox{\cmss Z\kern-.4em Z}}{\hbox{\cmss Z\kern-.4em Z}}
{\lower.9pt\hbox{\cmsss Z\kern-.4em Z}} {\lower1.2pt\hbox{\cmsss
Z\kern-.4em Z}}\else{\cmss Z\kern-.4em Z}\fi}
\def\IR{\relax{\rm I\kern-.18em R}}

\def\one{{\hbox{ 1\kern-.8mm l}}}

\newlength{\bredde}
\def\slash#1{\settowidth{\bredde}{$#1$}\ifmmode\,\raisebox{.15ex}{/}
\hspace*{-\bredde} #1\else$\,\raisebox{.15ex}{/}\hspace*{-\bredde}
#1$\fi}

\newsavebox{\zzzbar}
\sbox{\zzzbar}
  {\setlength{\unitlength}{0.9em}
  \begin{picture}(0.6,0.7)
  \thinlines
  \put(0,0){\line(1,0){0.6}}
  \put(0,0.75){\line(1,0){0.575}}
  \multiput(0,0)(0.0125,0.025){30}{\rule{0.3pt}{0.3pt}}
  \multiput(0.2,0)(0.0125,0.025){30}{\rule{0.3pt}{0.3pt}}
  \put(0,0.75){\line(0,-1){0.15}}
  \put(0.015,0.75){\line(0,-1){0.1}}
  \put(0.03,0.75){\line(0,-1){0.075}}
  \put(0.045,0.75){\line(0,-1){0.05}}
  \put(0.05,0.75){\line(0,-1){0.025}}
  \put(0.6,0){\line(0,1){0.15}}
  \put(0.585,0){\line(0,1){0.1}}
  \put(0.57,0){\line(0,1){0.075}}
  \put(0.555,0){\line(0,1){0.05}}
  \put(0.55,0){\line(0,1){0.025}}
  \end{picture}}

\newcommand{\ena}{\end{eqnarray}}
\newcommand{\beqa}{\begin{eqnarray}}
\newcommand{\eeqa}{\end{eqnarray}}

\newcommand{\half}{\frac{1}{2}}

\newfont{\goth}{ygoth.tfm scaled 1200}                   

 \numberwithin{equation}{section}

\def\1{{(1)}}
\def\2{{(2)}}
\def\3{{(3)}}

\def\1{{\bf 1}}

\begin{document}
\begin{titlepage}

\begin{center}

{April 9, 2012}
\hfill         \phantom{xxx}  EFI-12-4

\vskip 2 cm {\Large \bf The Conformal Anomaly of $M5$-Branes}
\vskip 1.25 cm {\bf  Travis Maxfield\footnote{maxfield@uchicago.edu} and Savdeep Sethi\footnote{sethi@uchicago.edu}}\non\\
\vskip 0.2 cm
 {\it Enrico Fermi Institute, University of Chicago, Chicago, IL 60637, USA}

\end{center}
\vskip 2 cm

\begin{abstract}
\baselineskip=18pt

We show that the conformal anomaly for $N$ $M5$-branes grows like $N^3$. The method we employ relates Coulomb branch interactions in six dimensions to interactions in four dimensions using supersymmetry. This leads to a relation between the six-dimensional conformal anomaly and the conformal anomaly of N=4 Yang-Mills. Along the way, we determine the structure of the four derivative interactions for the toroidally compactified $(2,0)$ theory, while encountering interesting novelties in the structure of the six derivative interactions.     

\end{abstract}

\end{titlepage}


\section{Introduction}
\label{intro}

One of the real surprises that emerged from the string duality web is the existence of six-dimensional interacting theories with $(2,0)$ supersymmetry~\cite{Witten:1995zh}. In type IIB string theory, these $(2,0)$ theories appear at $ADE$ singularities from $D3$-branes which wrap the vanishing cycles. The $A_N$ version of the theory is also found on $N+1$ coincident $M5$-branes. On circle reduction to five dimensions, the $(2,0)$ theory reduces to maximally supersymmetric Yang-Mills theory. 

A single $M5$-brane supports a six-dimensional two-form tensor with chiral field strength so it is natural to speculate that the $(2,0)$ theory is a generalization of Yang-Mills theory involving strongly interacting self-dual strings. We currently have no precise formulation of the theory, but the most striking known fact obtained from dual gravitational descriptions is that the number of degrees of freedom grows like $N^3$ for the $A_N$ theory~\cite{Klebanov:1996un, Henningson:1998gx, Bastianelli:2000hi}. The same cubic growth can be found from anomaly inflow for $M5$-branes coupled to $M$-theory~\cite{Harvey:1998bx, Yi:2001bz}. 

The generalization of gluon dynamics which could provide an $N^3$ growth is a fascinating question. Recent counts of $1/4$ BPS states on the Coulomb branch of both the $(2,0)$ theory and five-dimensional Yang-Mills theory see a suggestive $N^3$ growth, which is perhaps a hint about the nature of the fundamental degrees of freedom~\cite{Kim:2011mv, Bolognesi:2011rq, Bolognesi:2011nh}. 

Our goal in this work is to derive the cubic growth of the conformal anomaly from  field theory. We will study the Coulomb branch effective action for the $(2,0)$ theory. The Coulomb branch has a rich structure which includes couplings related to the $Spin(5)_R$ anomaly~\cite{Intriligator:2000eq}, and topological terms which modify the chiral string current~\cite{Ganor:1998ve}. Maximal supersymmetry implies rather powerful constraints on Coulomb branch interactions. These constraints were originally found in~\cite{Paban:1998ea, Paban:1998qy}, and largely applied to non-conformal models. We expect the combination of conformal invariance and maximal supersymmetry will provide yet more control over the Coulomb branch theory; this work can be viewed as a first foray in this direction.  

\subsection{The basic idea and an outline}

The conformal anomaly is computable on the Coulomb branch of the $(2,0)$ theory. It is captured by the six derivative interactions so our task is to understand those couplings. We will proceed in two steps: first, we determine the four derivative interactions of the $(2,0)$ theory compactified on a two-torus. This allows us to relate the four derivative interactions in six dimensions to four derivative interactions of N=4 Yang-Mills which are one-loop exact. We then show that the relevant six derivative interactions of the $(2,0)$ theory are determined by the four derivative interactions. This leads to the cubic growth of the conformal anomaly for $N$ $M5$-branes. In principle, the method we propose should provide a precise expression for the conformal anomaly of the $ADE$ $(2,0)$ theories. Determining the exact formula will require additional information about the Coulomb branch supersymmetry transformations and interactions up to six derivatives.   

In section two, we summarize our conventions and list the relevant supersymmetry transformations. Section three contains an analysis of the four derivative terms of the toroidally compactified $(2,0)$ theory for rank one. The four derivative couplings are harmonic on the moduli space which contains a circle. The solution is a sum of Bessel functions. It would be interesting to see whether the Fourier coefficients of these couplings  count BPS configurations in some suitable sense. 

Section four relates the six derivative interactions to the four derivative interactions again for rank one. In section five, we extend our analysis of four derivative couplings to rank $N$ along the lines described in~\cite{Sethi:1999qv}.  Finally in section six, we extract the growth of the conformal anomaly from our results. Among the outstanding remaining questions is understanding the extent to which all four and six derivative interactions are determined by the top form fermion couplings, or more generally, a complete understanding of the supersymmetry constraints.  
In terms of recent literature, work has focused on possibly defining the $(2,0)$ theory via gauge theory~\cite{Douglas:2010iu, Lambert:2010iw,  Ho:2011ni, Singh:2011id, Samtleben:2011fj, Chu:2011fd, Chu:2012um, Samtleben:2012mi},
studying amplitudes for the $(2,0)$ theory~\cite{Czech:2011dk}, circle compactifications~\cite{ Linander:2011jy, Gustavsson:2011ur}, the conformal anomaly of Wilson surfaces~\cite{Young:2011aa},  and connections with membrane theories~\cite{Berman:2008be, Berman:2009xd,  Berman:2009kj, Terashima:2010ji, Lambert:2011eg, Papageorgakis:2011xg}; for a review of prior developments, see~\cite{Berman:2007bv}.

\section{SUSY Transformations and Conventions}
\label{conventions}
\subsection{Effective actions and order counting}
In this work, we will analyze the low energy effective action which describes the Coulomb branch of superconformal field theories in $4$ and $6$ dimensions. We expand the Coulomb branch Lagrangian in a derivative expansion,  
\be
\mathcal{L} = \mathcal{L}_1 + \mathcal{L}_2 + \ldots,
\ee
with $\mathcal{L}_k$ including all terms of order $2k$. The order of an operator counts the number of derivatives plus half the number of fermions, and assigns order zero to moduli and gauge fields. The kinetic terms for both fermions and bosons are therefore of order $2$ and are contained in $\mathcal{L}_1$. With maximal supersymmetry, the theory described by $\mathcal{L}_1$ is a free theory with a flat metric. Singularities in the moduli space are the only loci where an interacting infrared theory can emerge. Away from such singularities, the leading interactions are captured by $\mathcal{L}_2$.  

In such an effective expansion, it is necessary to modify the supersymmetry transformations of the fields order by order to ensure that the Lagrangian is on-shell supersymmetric. We will make this explicit for the $4$ and $6$-dimensional theories in  the following discussion. 




\subsection{$\mathcal{N}=4$ SYM in $d=4$}
The supermultiplet in 4 dimensions for a theory with $16$ supersymmetries consists of a gauge field $A_{\mu}$, six scalars $\phi^i$ with $i=1,2,\ldots 6$, and four Weyl fermions $\psi^a_{\alpha}$ with $a=1,2,3,4$ and $\alpha =1,2$. The $\mathcal{N} = 4$ SUSY algebra includes a $Spin(6)\sim SU(4)$ $R$-symmetry under which the gauge field is neutral, the scalars transform in the $\mathbf{6}$, and the Weyl spinors $\psi^a_{\alpha}$ are in the $\mathbf{4}$  while their conjugates $\overline{\psi}_{a\dot{\alpha}}$ are in the $\mathbf{\overline{4}}$. Notice that we have 4 complex two-component spinors for a total of 16 real spinors as required for maximal supersymmetry.

The free particle supersymmetry variations of these fields can be determined by dimensional reduction from 10-dimensional N=1 Yang-Mills: 
\begin{align} \label{4dtrans}
\delta_{\epsilon} A_{\mu}& = i\epsilon^a\sigma_{\mu}\overline{\psi}_a + c.c. , \nonumber \\
\delta_{\epsilon} \phi^i &= i\epsilon^a\tau^i_{ab}\psi^b + c.c. , \\
\delta_{\epsilon} \psi^a &= F^{\mu \nu}\sigma_{\mu \nu}\epsilon^a + 2(\tau^i)^{ab}\sigma_{\mu}\partial^{\mu}\phi^i\bar{\epsilon}_b. \nonumber
\end{align}
The $\tau^i$ are elements of the reduced $Spin(6)$ Clifford algebra. These variations will be corrected at each order in the expansion of the action to ensure on-shell closure of the algebra. We will follow the notation of~\cite{Paban:1998ea, Paban:1998qy} and parametrize these corrections as follows,
\begin{align} \label{4dcorrectedtrans}
\delta_{\epsilon} A_{\mu}& = i\epsilon^a\sigma_{\mu}\overline{\psi}_a + \epsilon^a(K_{\mu})_a^{\phantom{a}b}\overline{\psi}_b + c.c. , \nonumber \\
\delta_{\epsilon} \phi^i &= i\epsilon^a\tau^i_{ab}\psi^b + \epsilon^aN^i_{ab}\psi^b + c.c. , \\
\delta_{\epsilon} \psi^a &= F^{\mu \nu}\sigma_{\mu \nu}\epsilon^a + 2(\tau^i)^{ab}\sigma_{\mu}\partial^{\mu}\phi^i\bar{\epsilon}_b  + M^a_b\epsilon^b + L^{ab}\overline{\epsilon}_b,\nonumber
\end{align}
where $K$, $L$, $M$, and $N$ are to be expanded order by order. Conformal invariance together with maximal supersymmetry ensures that the metric on the moduli space is flat and non-renormalized, so the first corrections encoded in $K$, $L$, $M$, and $N$ are induced by $\mathcal{L}_2$. These corrections are to be applied to the terms of $\mathcal{L}_1$ and will mix with the free variations acting on $\mathcal{L}_2$. Therefore, these first corrections must be of order 2, 3, 3, and 2 for $K$, $L$, $M$, and $N$, respectively.


\subsection{$\mathcal{N} = (2,0)$ in $d = 6$}
The $(2,0)$ tensor multiplet in six dimensions contains an antisymmetric tensor $B_{\mu \nu}$, five scalars $\Phi^I$, $I=1,\ldots,5$, and four symplectic-Majorana-Weyl fermions $\Psi^a$, $a=1,\ldots,4$. They transform under the $Spin(5)_R$ symmetry as $\mathbf{1}$, $\mathbf{5}$, and $\mathbf{4}$, respectively.

The tensor $B_{\mu \nu}$ in six dimensions naturally has mass dimension $2$ coming from the vanishing dimension of a Wilson surface operator. The scalars and fermions have canonical mass dimensions of $2$ and $5/2$, respectively. The tensor has an abelian gauge symmetry
\be
B_{\mu\nu} \rightarrow B_{\mu\nu} + \partial_{\mu}\Lambda_{\nu} - \partial_{\nu}\Lambda_{\mu}.
\ee
The exterior derivative of $B$ is a three form $H_{\mu \nu \lambda}=3\partial_{[\mu}B_{\nu \lambda ]}$, which is gauge invariant. A real three form in six-dimensional Minkowski space can be chosen to be self-dual or anti-self-dual. This choice correlates with the choice of handedness of the fermions in the supermultiplet. We will choose the fermions to have positive chirality, which implies the tensor is anti-self-dual:
\be \label{tensoreom}
H = - \ast H. 
\ee

The fermions are in the $(\mathbf{4},\mathbf{4})$ of $Spin(5,1)\times Spin(5)\sim Spin(5,1)\times Sp(2) $. These are separately pseudoreal representations, so the fermions can be made real, with 
\be
\overline{\Psi}_a  \equiv (\Psi_a)^T\Gamma^0 = M_{ab}(\Psi^b)^T\Gamma^0,
\ee
where $M_{ab}$ is the symplectic structure on $Sp(2)$; see, for example~\cite{Linander:2011jy}. The fermions have been chosen to have positive chirality: $\Gamma_V \Psi^a = \Psi^a$, where $\Gamma_V = \Gamma^0\ldots\Gamma^5$.

The SUSY transformations of the tensor multiplet component fields, including higher order corrections, are:
\begin{align} \label{6dtrans}
\delta_{\epsilon} B_{\mu \nu} = &-i \overline{\epsilon}_a\Gamma_{\mu \nu}\Psi^a +\overline{\epsilon}_a(K_{\mu\nu})^a_{\phantom{b}b}\Psi^b, \nonumber \\
\delta_{\epsilon} \Phi^I = &-i\overline{\epsilon}_a(\gamma^I)^a_{\phantom{x}b}\Psi^b + \overline{\epsilon}_a(N^I)^a_{\phantom{x}b}\Psi^b, \\
\delta_{\epsilon} \Psi^a = &\left(\frac{1}{12}H^{\mu \nu \lambda}\Gamma_{\mu \nu \lambda} \delta^a_{\phantom{x}b} + \Gamma^{\rho}\partial_{\rho}\Phi^I(\gamma^I)^a_{\phantom{x}b}\right)\epsilon^b + M^a_{\phantom{a}b}\epsilon^b. \nonumber
\end{align}
Here $\Gamma^{\mu}$ and $\gamma^I$ are the gamma matrices of $Spin(5,1)$ and $Spin(5)$, respectively, satisfying
\be
\acom{\Gamma^{\mu}}{\Gamma^{\nu}} = 2\eta^{\mu \nu}, \quad \acom{\gamma^I}{\gamma^J} = 2\delta^{I J}. \nonumber
\ee
Again, conformal invariance and maximal supersymmetry ensure the metric is flat, and the first corrections to $K$, $M$, and $N$ are of order 2, 3, and 2, respectively.

\subsection{$\mathcal{N} = (2,0)$ on $\mathbb{R}^{3,1}\times T^2$}
We will also consider the $(2,0)$ multiplet and its SUSY variations on a space-time of the form $\mathbb{R}^{3,1}\times T^2$, where we take the torus to be in the $(x^4, x^5)$ plane. For this analysis, we can restrict to a rectangular torus; there are interesting extensions to twisted cases which generate non-simply-laced $4$-dimensional gauge groups where including a $\theta$-angle is important, but we will not consider those cases here; see, for example,~\cite{Keurentjes:2002dc, Tachikawa:2011ch}. The Lorentz $Spin(5,1)$ symmetry is spontaneously broken to $Spin(3,1)$. For fields with zero momentum along the compact directions, the field content reduces to that of the 4-dimensional $\mathcal{N} = 4$ multiplet, except for the presence of one periodic scalar. This scalar, a Wilson surface variable, results from integration of  $B_{45}$ over the torus. The presence of this periodic scalar breaks the $Spin(6)$ symmetry of the $4$-dimensional theory to $Spin(5)$ acting on the non-compact scalars.

When dealing with the $(2,0)$ theory on a torus, we will rescale the $6$-dimensional fields to have the scaling dimensions of the $4$-dimensional theory. The rescaled scalars and fermions will be denoted by lower case Greek letters, $\phi$ and $\psi$, as in $4$ dimensions, with
\be
\phi \sim \sqrt{V_T}\Phi, \quad \ \psi \sim \sqrt{V_T}\Psi,
\ee
where $V_T = R_4 R_5$ is the volume of $T^2$. Since the scalars only have a $Spin(5)$ symmetry for non-zero $\sqrt{V_T}$, we will still  use upper case Latin letters to denote the $Spin(5)$ vector indices of $\phi^I$, $I=1,2,\ldots 5$. 

The periodic scalar will be denoted $\phi^6$. With  $4$-dimensional canonical scaling dimensions, this field has a period given by the inverse of the square root of the volume of the torus: 
\be
\phi^6 \sim \phi^6 + \frac{2\pi}{\sqrt{V_T}}.
\ee
It is important to stress that this choice of normalization gives scalar kinetic terms of the form, 
\be \int d^4x \, \partial \phi \partial \phi, \ee
without an overall ${1\over (g_4)^2}$ factor. This factor can be restored after making a choice of gauge coupling; either $(g_4)^2 = R_4/R_5$, or the S-dual choice $R_5/R_4$.  

The free particle supersymmetry variations enjoy the full $Spin(3,1)\times Spin(6)$ symmetry of the $4$-dimensional theory since their form is independent of the topology of the moduli space. For a single multiplet, the moduli space is locally $\mathbb{R}^5\times S^1$. The supersymmetry variations are therefore identical to those of the $4$-dimensional theory appearing in~\C{4dtrans}. This is certainly not true of the higher momentum terms! As we will see, those couplings can detect the circle in the moduli space.  

The fermions, originally in the $(\mathbf{4},\mathbf{4})$ of $Spin(5,1)\times Spin(5)$, decompose into $(\mathbf{2},\mathbf{4}) \oplus (\mathbf{\overline{2}},\mathbf{4})$ under the residual $Spin(3,1)\times Spin(5)$ symmetry. Since the free particle supersymmetry variations respect the full $4$-dimensional symmetry group, this representation can instead be viewed as the $(\mathbf{2},\mathbf{4}) \oplus (\mathbf{\overline{2}},\mathbf{\overline{4}})$ of $Spin(3,1)\times Spin(6)$. We will at times find it more convenient to use one representation over the other. It should be clear which representation we are considering by the choice of gamma matrices for the $R$-symmetry: $\tau^i$ denotes the $Spin(6)$ representation, while $\gamma^I$ denotes the $Spin(5)$ representation.

\section{Eight Fermion Terms for Rank One}\label{determiningeight}

We would like to examine the constraints imposed by sixteen supersymmetries on the four derivative terms in 6 and 4 dimensions, and the solution that interpolates between them. We begin by restricting to theories with rank one: the low-energy theory is either gauge symmetry spontaneously broken to $U(1)$, or the $(2,0)$ theory broken to a single free $(2,0)$ multiplet. Our analysis will parallel that of~\cite{Paban:1998qy,Paban:1998mp}. In each dimension, we focus our analysis on terms in $\mathcal{L}_2$ with eight fermions and no derivatives. These terms are the ``top forms'' of the action at that order, and the simplest couplings to control. 

\subsection{$\mathcal{N} = (2,0)$ in $d=6$} \label{eightfortwozero}
In six dimensions, the interacting rank one $(2,0)$ theory has a moduli space that is  $\mathbb{R}^5/\Z_2$. There is a single scale in the problem which is the vacuum expectation value (VEV), denoted $V$, for the scalar field where $V^2 = \sum_I \langle \Phi^I \Phi^I \rangle$. The $k=1$ action describes the action for a free tensor multiplet. Because of the chirality constraint on the tensor multiplet, we should really describe the theory in terms of equations of motion rather than an action. However, it is more intuitive to think in terms of an action. Since we are largely concerned with the moduli-dependence of scalar and fermion couplings, using the terminology of an action rather than equations of motion will not alter any physical conclusions, as long as we impose the first order equation of motion on the chiral tensor~\C{tensoreom}.  




The first interaction terms in the action appear at order $k=2$. In a standard field theory, these terms are expected to be produced from the integrating out of modes with masses $m^2 \sim V$. For the $(2,0)$ theory, what should be integrated out is mysterious yet the Coulomb branch effective action is nevertheless well-defined. 
In particular, consider an eight fermion term in $\mathcal{L}_2$. Since it cannot contain any derivatives, it must be composed solely of moduli and fermions, schematically
\be\label{eightfermion}
\mathcal{L}_2 \supset f^{(8)}(\Phi)\Psi^8.
\ee
For the action to be supersymmetric, the variation of this term must either vanish or cancel with the variation of some other term in the action.  Let us focus on a particular variation of the eight fermion term, specifically the supersymmetry variation acting on the moduli:
\be
\delta_{\epsilon}\mathcal{L}_2 \supset (\delta_{\epsilon}\Phi)\frac{\partial}{\partial \Phi}\left(f^{(8)}(\Phi)\Psi^8\right) + (\delta_{\epsilon}\Psi)\frac{\partial}{\partial \Psi}\left(f^{(8)}(\Phi)\Psi^8\right).
\ee
The first term on the right contains $9$ fermions. No other term in the SUSY variation of $\mathcal{L}_2$ contains $9$ fermions and no derivatives. However, it could be possible that the corrected supersymmetry variations of~\C{6dtrans} produce such a term when acting on $\mathcal{L}_1$. At this point, the flat moduli space metric provides a nice simplification. Because the metric is flat, $\mathcal{L}_1$ contains at most $2$ fermion couplings and no $4$ fermion interactions which would couple to curvature. Applying the corrected supersymmetry transformations (which are of order $3$ at most) cannot produce a $9$ fermion term. We therefore find, as in~\cite{Paban:1998ea}, that the $9$ fermion term must vanish on its own:
\be \label{6dfullconstraint}
(\gamma^I)^a_{\phantom{x}b}\Psi^b_{\alpha}\frac{\partial}{\partial \Phi^I} \left(f^{(8)}(\Phi)\Psi^8\right) = 0.
\ee

To make this more precise, we need to classify the possible structures that can appear in the action. Specifically, which representations occur in the antisymmetric product of  eight fermions. Since there are no derivatives in this term, the eight fermion product must be Lorentz invariant on its own. However, it does not need to be $Spin(5)$ invariant since we can contract any $Spin(5)$ vector indices with scalars to form an invariant coupling. Hence, we expect the $8$ fermion term to take the form,
\be
f^{(8)}\Psi^8 \sim \sum_k f^{(8)}_k(r)\Phi^{I_1}\Phi^{I_2}\ldots\Phi^{I_k}T^{I_1I_2\ldots I_k},
\ee
with $T^{I_1I_2\ldots I_k}$ a completely symmetric tensor appearing in the product of eight fermions. We have also introduced the $Spin(5)$ invariant $r^2 = \sum_I(\Phi^I)^2$. 

We can determine which $T^{I_1\ldots I_k}$ are possible from contractions of the basic bilinears appearing in the antisymmetric product of two fermions in the $(\mathbf{4},\mathbf{4})$ of $Spin(5,1)\times Spin(5)$:
\be
\overline{\Psi}\Gamma^{\mu}\gamma^{IJ}\Psi, \quad \overline{\Psi}\Gamma^{\mu\nu\lambda}_{+}\gamma^{I}\Psi, \quad \overline{\Psi}\Gamma^{\mu\nu\lambda}_{+}\Psi,
\ee
where $\Gamma^{\mu}$ are the six-dimensional reduced gamma matrices, $\gamma^I$ are the gamma matrices of $Spin(5)$ and $\Gamma^{\mu\nu\lambda}_{+} =\Gamma^{[\mu\nu\lambda]}_{+} $ are the self-dual $3$-index gamma matrices. We build a general eight fermion term from some contraction of four of these bilinears. Since the scalars which multiply the eight fermion term must be fully symmetric, we see that we can have no more than one scalar associated with each bilinear. Therefore, we are limited to $T^{I_1\ldots I_k}$ with $0\leq k\leq 4$. It is possible to restrict the set of representations further by the use of various Fierz identities. Instead, we will perform the same task using representation theory.\footnote{We found the group-theoretic program LiE useful for this task~\cite{LIE}}. In the decomposition of $\wedge^8 (\mathbf{4},\mathbf{4})$, each $T^{I_1\ldots I_k}$ for $0\leq k\leq 4$ occurs precisely once, where we now work in the basis where each $T^{I_1\ldots I_k}$ is a traceless irreducible representation of $Spin(5)$.

We will find that supersymmetry and conformal invariance restrict the allowed representations to only the traceless, symmetric four index tensor, $T^{IJKL}$. To see this, we act on~\C{6dfullconstraint} with the operator $(\gamma^J)^c_{\phantom{x}a} \partial_{\Psi^c_{\alpha}} \partial_{\Phi^J}$. After summing over $a$ and $\alpha$, we obtain a weaker supersymmetry condition:
\be \label{6dlaplaceconstraint}
\sum_I \left(\frac{\partial}{\partial\Phi^I}\right)^2 \left(f^{(8)}(\Phi)\Psi^8\right) \equiv \Delta \left(f^{(8)}(\Phi)\Psi^8\right) = 0.
\ee

This condition, unlike the original, does not mix fermion structures with different numbers of scalar contractions which are needed in $f^{(8)}$ to build a traceless representation. Consider the Laplace equation on the four scalar structure first:
\be
\Delta\left(f^{(8)}_4(r)\Phi^I\Phi^J\Phi^K\Phi^LT^{IJKL}\right) =0 \quad \Rightarrow \quad \frac{d^2}{dr^2}f^{(8)}_4(r) + \frac{12}{r}\frac{d}{dr}f^{(8)}_4(r) = 0,
\ee
the solution of which is
\be
f^{(8)}_4(r) = \frac{c_{(6)}}{r^{11}} + d
\ee
where $c_{(6)}$ and $d$ are numbers determined (in principle) through knowledge of the interacting $(2,0)$ theory. Note that since there are no couplings for the $(2,0)$ theory, they have no parametric dependence. We also must impose  conformal invariance as a constraint. As noted before, the fermions and scalars of the $(2,0)$ multiplet have mass dimensions of $5/2$ and 2, respectively. From this, we see that the $c_{(6)}$ solution is conformally invariant, while the constant is not; therefore, $d$ = 0. The constant solution is also unphysical since it is not suppressed as $r\rightarrow\infty$.  In subsequent discussion, we will just ignore the constant solutions.

Next consider the structure with $k$ scalars. The Laplace equation reads:
\be
\frac{d^2}{dr^2}f^{(8)}_k(r) + \frac{4 + 2k}{r}\frac{d}{dr}f^{(8)}_k(r) = 0 \quad \Rightarrow \quad f^{(8)}_k(r) = \frac{c_k}{r^{3+2k}}.
\ee
However, conformal invariance demands that $f^{(8)}_k(r)$ be a homogeneous function of degree $-7-k$, since there are no dimensionful parameters. This uniquely fixes $k=4$. We have therefore determined the eight fermion term of the four derivative action up to a single constant $c_{(6)}$.

With the eight fermion terms determined, the only remaining question is the extent to which these couplings determine all the other couplings in $\mathcal{L}_2$. For maximally supersymmetric Yang-Mills in $0+1$-dimensions, it is known that all other couplings in $\mathcal{L}_2$ are determined by the eight fermion terms~\cite{Hyun:1999hf, Kazama:2002jm, Sethi:2004np}. It is basically the statement that there is a single super-invariant of order $4$ which terminates at eight fermions. The same is true in $3+1$-dimensions~\cite{Buchbinder:2001xy, Belyaev:2011wa, Belyaev:2011dg, Buchbinder:2011zu}. We fully expect this to be true in six dimensions as well, but it remains to be demonstrated. 

\subsection{$\mathcal{N} = 4$ SYM in $d=4$} \label{n=4}
The same analysis can be performed in four dimensions. Again, we find that supersymmetry demands the eight fermion term have moduli dependence that is harmonic:
\be
\Delta\left(f^{(8)}(\phi)\psi^8\right)=0.
\ee
We are again restricted to considering eight fermion structures with fewer than 5 symmetric, traceless $Spin(6)$ vector indices. This time, a representation theory calculation with LiE restricts,
\be
\psi^8 \supset T_{m_0}, \, \ T^{IJ}_{m_2}, \, \ T^{IJKL}_{m_4}, 
\ee
where the $m_k$ label the number of linearly independent such terms appearing in the product of eight fermions. In this case, $m_0=m_2=3$ and $m_4=1$.

The $k$ scalar constraint is
\be
\frac{d^2}{dr^2}f^{(8)}_k(r) + \frac{5 + 2k}{r}\frac{d}{dr}f^{(8)}_k(r) = 0 \quad \Rightarrow \quad f^{(8)}_k(r) = \frac{c_k}{r^{3+2k}}.
\ee
Conformal invariance requires that $f^{(8)}_k(r)$ be homogeneous of degree $-8-k$, requiring  $k=4$ as before. The constant appearing in this solution will be denoted $c_{(4)}$ in analogy with six dimensions. Furthermore, since the four derivative action of $\mathcal{N}=4$ theories in four dimensions is known to be one-loop exact~\cite{Dine:1997nq}, this coefficient will have no dependence on  the Yang-Mills coupling $g^2$. It is satisfying to note dependence on only one constant. Had there been two or more linearly independent four scalar structures, each with independent coefficients, we would have a potential disagreement with many known results. 

\subsection{$\mathcal{N}=(2,0)$ on $T^2$} \label{rank1torus}
Next, we will show that $c_{(6)}$ and $c_{(4)}$ are simply related. We will do so by examining the effective action of the six dimensional theory on the space-time $\mathbb{R}^{3,1}\times T^2$ for a torus of finite size. It is believed that in the limit of vanishing torus, this is the $4$-dimensional $\mathcal{N}=4$ theory of the previous section. The eight fermion terms on the torus should interpolate between $4$ and $6$ dimensions, agreeing with the $4$ $(6)$ dimensional result in the $V_T\to 0 \, (\infty)$ limit. We will see that this matching actually specifies $c_{(6)}$ in terms of $c_{(4)}$.


With the conventions of the previous section, the basic fermion bilinears from which the eight fermion terms can be built are
\begin{align}
\psi & \gamma^{IJ}\psi \nonumber \\
\psi\sigma^{\mu}\overline{\psi}, \quad \psi\sigma^{\mu} & \gamma^I\overline{\psi}, \quad \psi\sigma^{\mu}\gamma^{IJ}\overline{\psi} \\
\psi \sigma^{\mu\nu} \psi, \, \ & \, \ \psi\sigma^{\mu\nu}\gamma^I\psi, \nonumber
\end{align}
where complex conjugates should also be included where applicable. Wedging these structures together yields the following eight fermion structures:
\be
\psi^8 \supset T_{m_0}, \quad T^I_{m_1}, \quad T^{IJ}_{m_2}, \quad T^{IJK}_{m_3}, \quad T^{IJKL}_{m_4},
\ee
where now $m_0=7$, $m_1=m_2=6$, $m_3=2$, and $m_4=1$. The functions $f^{(8)}_k(r)$ that multiply these structures now will depend on all the scalars where $\phi^I$ must appear in the combination $r=\sqrt{\phi^I\phi^I}$ and $f^{(8)}_k$ must respect the periodicity of $\phi^6$. We will Fourier expand in this variable; the function which multiplies the $a^{\rm th}$ structure with $k$ scalars then takes the form:
\be
f_k^{(8),a}(r,\phi^6) = \sum_n f_{k,n}^{(8),a}(r)e^{i n \phi^6\sqrt{V_t}}.
\ee

Again consider a supersymmetry variation of the general eight fermion term $f^{(8)}\psi^8$. We find that the piece with nine fermions must vanish,
\be \label{torusconstraint}
\tau^i_{ab} \psi^{\alpha b} \frac{\partial}{\partial \phi^i} \Big( f^{(8)}\psi^8 \Big) = (\tau^i)^{ab} \overline{\psi}^{\dot{\alpha}}_{b} \frac{\partial}{\partial \phi^i} \Big( f^{(8)}\psi^8 \Big) = 0,
\ee
where the two equations are Hermitian conjugates. Applying the operator
\be
(\tau^j)^{ca}\frac{\partial}{\partial \psi^{\alpha c}}\frac{\partial}{\partial \phi^j}
\ee
to the right hand side yields a Laplacian with the periodic scalar $\phi^6$ included:
\be
\Delta\Big(f^{(8)}(r,\phi^6)\psi^8\Big)=0.
\ee
As before, this leads to a homogeneous equation for the function $f^{(8),a}_k(r,\phi^6)$ which can be separated into its Fourier components:
\be
\frac{d^2}{dr^2}f^{(8),a}_{k,n} +\frac{4+2k}{r}\frac{d}{dr}f^{(8),a}_{k,n} - n^2V_Tf^{(8),a}_{k,n} = 0.
\ee

\subsubsection{Zero-mode solutions}
Focusing on the zero Fourier mode gives,
\be \label{torussol1}
f^{(8),a}_{k,0} = V_T^{(5-k)/2}\frac{c^a_{k,0}}{r^{3+2k}},
\ee
where the power of $V_T$ has been chosen so that $c_{k,0}$ is dimensionless. Unlike the non-compact cases, we can no longer rule out solutions based on conformal invariance since we have the freedom to insert powers of $V_T$ for dimensional agreement.


\subsubsection{Higher Fourier modes}
Now we would like to examine the non-zero Fourier components of the functions $f^{(8),a}_k(r,\phi^6)$. Consider the unique four scalar equation:
\be \label{rank1bessel}
\frac{d^2}{dr^2} f^{(8)}_{4,n} +\frac{2\nu + 1}{r}\frac{d}{dr}f^{(8)}_{4,n} - n^2V_Tf^{(8)}_{4,n} = 0, \quad \nu = 11/2.
\ee
The solution to this equation that is regular at $r=\infty$ is given by a modified Bessel function of the second kind:
\be \label{highermodesusyform}
f^{(8)}_{4,n} = c_{4, n} |n|^{1/2}V_T^6\Big(V_T^{1/2}r\Big)^{-\nu}K_{\nu}(|n| rV_T^{1/2}),
\ee
where factors of $V_T$ have been chosen so $c_{4,n}$ is dimensionless.

In principle we could solve the set of equations at each $n$ in terms of these Bessel function solutions, each with a seemingly independent coefficient. However, we expect that in the limit of $V_T \to 0$, we should recover the $4$-dimensional eight fermion term from some combination of each of the terms on the torus. Here it is necessary that there are contributions from each of the $k$ scalar structures, which should recombine into a $4$ scalar structure of $Spin(6)$. 

This suggests that the coefficients of the different $k$ scalar structures are actually not independent and must be related. In fact, we will find that the more powerful SUSY constraint of~\C{torusconstraint} demonstrates this. First note that the Bessel function $K_{\nu}$ has the following expansion  for  half-integer $\nu$:
\be
K_{m+1/2}(z) = (\pi/2z)^{1/2}e^{-z}\sum_{n=0}^m(2z)^{-n}\frac{\Gamma(m+n+1)}{n!\Gamma(m-n+1)}.
\ee
Our solution to the four scalar constraint equation can then be expanded as follows, 
\be
f^{(8)}_{4,n}e^{in\phi^6\sqrt{V_T}} =  c_{4,n} |n|^{1/2}V_T^6\left(\frac{\pi}{2|n|}\right)^{1/2}e^{-|n|r\sqrt{V_T} + in\phi^6\sqrt{V_T}}\left( \left(V_T^{1/2}r\right)^{-6} + \ldots \right),
\ee
where the extra terms are more singular as $r\to 0$. If we include the norm of the four scalars, the least singular term behaves like
\be
\sim   {1\over r^2}e^{-|n|r\sqrt{V_T}}.
\ee

For a term with $k$ scalars, each solution will also have a least singular norm of $r^{-2}e^{-|n|r\sqrt{V_T}}$. Note that we are not relying on any asymptotic expansion for either small or large $r$ since we have an exact solution for each Fourier mode. The constraint~\C{torusconstraint}\ contains a piece that only involves terms with this singular behavior which we can isolate,
\be \label{orderconstraint}
\left(\frac{1}{r}\phi^I\tau^I_{ab} - i\text{ sign}(n)\tau^6_{ab}\right)\psi^b_{\alpha}h =\left(\frac{1}{r}\phi^I(\tau^I)^{ab} - i\text{ sign}(n)(\tau^6)^{ab}\right)\overline{\psi}_{b\dot{\alpha}}h=0,
\ee
where $I=1,\ldots,5$ and $h$ denotes the eight fermion terms of order $r^{-2}e^{-|n|r\sqrt{V_T}}$. Let us define $\tau^r = \frac{1}{r}\phi^I\tau^I$ and apply the operator $(\tau^r)^{ca}\partial_{\psi^c_{\alpha}}$ to the leftmost constraint (and its conjugate to the other constraint). Summing on repeated indices and adding the two constraints gives:
\be
\left(\left[\delta^c_{\phantom{x}b} - i\text{ sign}(n)(\tau^{r6})^c_{\phantom{x}b}\right]\frac{\partial}{\partial \psi^c_{\alpha}}\psi^b_{\alpha} + \left[\delta_c^{\phantom{x}b} - i\text{ sign}(n)(\tau^{r6})_c^{\phantom{x}b}\right]\frac{\partial}{\partial \overline{\psi}_c^{\dot{\alpha}}}\overline{\psi}_b^{\dot{\alpha}}\right)h=0.
\ee
Acting on an eight fermion term, the operator
\be
\left(\frac{\partial}{\partial \psi^a_{\alpha}}\psi^a_{\alpha} +\frac{\partial}{\partial \overline{\psi}_a^{\dot{\alpha}}}\overline{\psi}_a^{\dot{\alpha}}\right)h=8h,
\ee
so we have
\be \label{eigenconstraint}
\left(8 - i\text{ sign}(n)\left((\tau^{r6})^c_{\phantom{x}b}\frac{\partial}{\partial \psi^c_{\alpha}}\psi^b_{\alpha} + (\tau^{r6})_c^{\phantom{x}b}\frac{\partial}{\partial \overline{\psi}_c^{\dot{\alpha}}}\overline{\psi}_b^{\dot{\alpha}}\right)\right)h=0.
\ee
Using the same arguments as in~\cite{Paban:1998mp}, any Lorentz-invariant eight fermion term must have an equal number of $\psi^a_1$ and $\psi^a_2$ as well as an equal number of $\overline{\psi}_{\dot{1}a}$ and $\overline{\psi}_{\dot{2}a}$. Additionally, because of the Pauli matrix identities, 
\be
\sigma^{\mu}_{\alpha\dot{\alpha}}\sigma_{\mu}^{\beta\dot{\beta}} = 2\delta_{\alpha}^{\phantom{\alpha}\beta}\delta_{\dot{\alpha}}^{\phantom{\alpha}\dot{\beta}}, \,\, \text{etc.},
\ee
we can construct any eight fermion term from combinations of 
\begin{align} \label{eightfermterms}
& \psi_1\tau^{ijk}\psi_2\psi_1\tau^{lmn}\psi_2\psi_1\tau^{opq}\psi_2\psi_1\tau^{stu}\psi_2 + c.c, \nonumber \\
& \psi_1\tau^{ijk}\psi_2\psi_1\tau^{lmn}\psi_2\psi_1\tau^{opq}\psi_2\overline{\psi}_{\dot{1}}\tau^{stu}\overline{\psi}_{\dot{2}} + c.c, \\
& \psi_1\tau^{ijk}\psi_2\psi_1\tau^{lmn}\psi_2\overline{\psi}_{\dot{1}}\tau^{opq}\overline{\psi}_{\dot{2}}\overline{\psi}_{\dot{1}}\tau^{stu}\overline{\psi}_{\dot{2}}, \nonumber
\end{align}
where $i,j,\dots, u = 1,\ldots,6, r$.

The operator appearing in~\C{eigenconstraint}\ acts on one of the basic bilinears of~\C{eightfermterms}\ as follows,
\be\label{givescomm}
i\text{ sign}(n)\left((\tau^{r6})^c_{\phantom{x}b}\frac{\partial}{\partial \psi^c_{\alpha}}\psi^b_{\alpha} + (\tau^{r6})_c^{\phantom{x}b}\frac{\partial}{\partial \overline{\psi}_c^{\dot{\alpha}}}\overline{\psi}_b^{\dot{\alpha}}\right)\psi_1\tau^{ijk}\psi_2 = \psi_1\left(i\text{ sign}(n)[\tau^{r6},\tau^{ijk}]\right)\psi_2,
\ee
with a similar result for the conjugate bilinear. We evaluate the commutator appearing on the right hand side of~\C{givescomm}, 
\be
[\tau^{r6},\tau^{ijk}] = 6\left(\tau^r\delta^{6[i}\tau^{jk]} - \delta^{r[i}\tau^{jk]}\tau^6\right).
\ee
Using this result, we see that the operator appearing in~\C{eigenconstraint}\ has eigenvalues of $2,0,-2$ acting on a bilinear, where the eigenvalue of $2$ corresponds to the bilinear
\be \label{eigenbilinear}
\psi_1\left(\tau^{ijr}-i\text{ sign}(n)\tau^{ij6}\right)\psi_2, \quad i,j\neq r,6.
\ee
Therefore to satisfy the constraint equation, we must replace every bilinear in~\C{eightfermterms}\ with the above bilinear (or its conjugate) and contract the remaining indices with each other to form a $Spin(5)$ invariant term. It would appear that we are left with at least three independent eight fermion terms in $h$, each with an undetermined, independent coefficient. However, again following~\cite{Paban:1998mp}, we can rule out all but one eight fermion term.

Acting on the equation~\C{orderconstraint} with $\tau^r$ yields for positive $n$:
\be
\left(I-i\tau^{r6}\right)^a_{\phantom{a}b}\psi^b_{\alpha}h=\left(I-i\tau^{r6}\right)_a^{\phantom{a}b}\overline{\psi}_{b\dot{\alpha}}h=0.
\ee
The operators $\left(I\pm i\tau^{r6}\right)/2$ are complementary projection operators with trace equal to 2 and eigenvalues $1$ or $0$. This implies that each operator has two non-zero eigenvalues and two zero eigenvalues. Therefore, we can separate both $\psi$ and $\overline{\psi}$ separately into two classes: those that vanish when operated on with $\left(I - i\tau^{r6}\right)/2$ and those that are invariant. That $h$ vanishes when multiplied by each of the latter type of fermion implies that it is built from this type of fermion. Since there are eight of these (counting barred and unbarred), we have completely determined $h$ which is composed of the following fermions:
\be\label{structureh}
h \sim \psi_1^{(-,1)}\psi_2^{(-,1)}\psi_1^{(-,2)}\psi_2^{(-,2)}\overline{\psi}_{\dot{1}}^{(-,1)}\overline{\psi}_{\dot{2}}^{(-,1)}\overline{\psi}_{\dot{1}}^{(-,2)}\overline{\psi}_{\dot{2}}^{(-,2)}.
\ee
The $\psi^{(-,i)}$ are the non-vanishing components of $\left(I - i\tau^{r6}\right)\psi^A$, and similarly for the barred fermions. 
There is additionally a $Spin(5)$ structure attached to this term.  

Let us write this eight fermion coupling in a more convenient way. Such a term will arise from some combination of the contractions of the third term appearing in~\C{eightfermterms}\ in which each bilinear is replaced with the eigenbilinear of~\C{eigenbilinear}. Let us use $B^{ij}$ to denote $\left(\tau^{ijr}-i\text{ sign}(n)\tau^{ij6}\right)$ and write the combination of contractions as follows,
\begin{align} \label{defS}
S \equiv a_1 & \psi_1B^{ij}\psi_2\psi_1B^{ij}\psi_2\overline{\psi}_{\dot{1}}B^{kl}\overline{\psi}_{\dot{2}}\overline{\psi}_{\dot{1}}B^{kl}\overline{\psi}_{\dot{2}} \nonumber\\
+ a_2 & \psi_1B^{ij}\psi_2\psi_1B^{jk}\psi_2\overline{\psi}_{\dot{1}}B^{kl}\overline{\psi}_{\dot{2}}\overline{\psi}_{\dot{1}}B^{li}\overline{\psi}_{\dot{2}} \\
+ a_3 & \psi_1B^{ij}\psi_2\psi_1B^{kl}\psi_2\overline{\psi}_{\dot{1}}B^{ij}\overline{\psi}_{\dot{2}}\overline{\psi}_{\dot{1}}B^{kl}\overline{\psi}_{\dot{2}} \nonumber \\
+ a_4& \psi_1B^{ij}\psi_2\psi_1B^{kl}\psi_2\overline{\psi}_{\dot{1}}B^{jk}\overline{\psi}_{\dot{2}}\overline{\psi}_{\dot{1}}B^{li}\overline{\psi}_{\dot{2}}, \nonumber
\end{align}
where the relative values of $a_1$, $a_2$, $a_3$, $a_4$ are determined by the constraint that $S$ has the fermion content required for $h$~\C{structureh}. Note that $S$ contains all of the scalar structures so we have determined $h$ for any positive $n$,\footnote{For negative $n$ we must construct our eight fermion term from the other type of fermion $\psi^{(+,i)}$, which are non-vanishing eigenvectors of the $I+i\tau^{r6}$ projector.} 
\be \label{hfermterm}
h=c_n n^{1/2}V_T^6\left(\frac{1}{V_T^{1/2}r}\right)^{11/2}r^4e^{-nr\sqrt{V_T}}\left(\frac{\pi}{2nrV_T^{1/2}}\right)^{1/2}e^{in\phi^6\sqrt{V_T}}S.
\ee
We can extract any $k$ scalar term from $S$. The factor of
\be e^{-nr\sqrt{V_T}}\left(\frac{\pi}{2nrV_T^{1/2}}\right)^{1/2} \ee
in $h$ can then  be replaced with the Bessel function $K_{3/2 + k}(nrV_T^{1/2})$ to give the homogeneous $k$ scalar solution. Thus the solution to the constraint equation for each $n$ and for any number of scalars is completely determined up to the choice of the constant $c_{n}$. It now remains to determine these constants.

\subsubsection{Determining the coefficients $c_n$}
To determine $c_n$, we will use our knowledge that the coefficient functions should reduce to those of $\mathcal{N}=4$ SYM in the vanishing torus limit $V_T\to0$. Let us consider the $4$-dimensional four scalar term,
\be
f^{(8)}_4  = \frac{c_{(4)}}{(r^2 + (\phi^6)^2)^6}\phi^{i} \phi^{j} \phi^{k}\phi^{l}T^{ijkl},
\ee
where we have separated out $\phi^6$ from the other scalars. If we choose a term that does not involve $\phi^6$  contracted with $T$, it must descend from the four scalar term on the torus as $V_T\to0$. 

To extract the $c_n$, we will first make the coefficient of the $4$-dimensional eight fermion term (discussed in section~\ref{n=4}) periodic in $\phi^6$ by summing over translations. We denote this periodic coefficient function by $F_p$: 
\bea
F_p &=& \sum_n\frac{c_{(4)}}{(r^2+(\phi^6+nV_T^{-1/2})^2)^6} = \sum_n\frac{c_{(4)}V_T^6}{((V_T^{1/2}r)^2+(V_T^{1/2}\phi^6+n)^2)^6}, \cr  &\equiv & \sum_nF(V_T^{1/2}\phi^6+n).
\eea
This is equal by Poisson resummation to
\be
F_p = \sum_n\hat{F}(n)e^{in\phi^6\sqrt{V_T}},
\ee
where $\hat{F}$ is the Fourier transform of $F$. We see that the $n^{\mathrm{th}}$ Fourier component in the series of $F_p$ is
\be\label{Fp}
F_p^{(n)} = c_{(4)}V_T^6\frac{2\pi^{1/2}}{5!}\left(\frac{n}{2rV_T^{1/2}}\right)^{11/2}r^4K_{11/2}(nrV_T^{1/2}).
\ee
This agrees precisely with the form required by supersymmetry~\C{highermodesusyform}. This need not have been the case, but it seems to be generally true for order $4$ interactions probably because the supersymmetry constraints imply a harmonic equation without source. For higher order interactions, we do not expect this to be true. 

In this case, however, we can compare~\C{Fp}\ to the component extracted from~\C{hfermterm}\ which yields:
\be \label{fouriercoeff}
c_n = c_{(4)}\left(\frac{n^5\sqrt{\pi}}{2^{9/2}5!}\right), \quad n\neq0.
\ee
Now that we have determined all of the non-zero Fourier modes of the eight fermion term on $\mathbb{R}^{3,1}\times T^2$, we can examine the $n=0$ modes again. In particular, we will make the claim that $c^a_{k,0} =0$ for $k\neq 4$, while $c_{4,0}\propto c_{(4)}$. To show this last statement, we replace $c_n$ in~\C{hfermterm} with~\C{fouriercoeff}. To isolate the four scalar structure, we then make the replacement prescribed just after~\C{hfermterm}. Taking the $n\to 0$ limit yields:
\be
c_{4,0}=c_{(4)}\left(\frac{63\pi}{2^8}\right).
\ee
Next, we isolate the $k<4$ scalar structures and again take the $n\to 0$ limit, this time yielding $c^a_{k,0}=0$.

We have yet to show that $c_{(4)}$ is related to $c_{(6)}$, as was our stated intent. However, this is clear from the form of the $4$ scalar zero mode solution. After rescaling the fields to have their $6$-dimensional dimensions, 
\be
\phi \sim \sqrt{V_T}\Phi, \, \ \psi \sim \sqrt{V_T}\Psi, \nonumber
\ee
we see that the four scalar structure on the torus is simply the dimensional reduction of the four scalar eight fermion term in six dimensions. Therefore, we have found that
\be
c_{(4)} \propto c_{4,0} \propto c_{(6)}.
\ee

\section{Six Derivative Couplings in Six Dimensions}

For Yang-Mills theories with maximal supersymmetry in dimension four or lower, the top form for the six derivative interactions is a $12$ fermion coupling. Surprisingly, this is not the case for the $(2,0)$ theory with rank one. If one attempts to build a suitable Lorentz invariant combination of  
twelve fermions which transform in the $(\mathbf{4},\mathbf{4})$ of $Spin(5,1)\times Spin(5)$, where suitable means contractable with scalars to give a $Spin(5)$ invariant, one finds that none exist. Rather, checking the representations which appear in the wedge product of  $(\mathbf{4},\mathbf{4})$, we find that the first potential top form actually consists of $1$ derivative and $10$ fermions. This is a consequence of the chiral nature of the theory. 

Let us focus on top forms which contain $1$ three-form field strength $H$ and $10$ fermions, with some number of scalars. The possibilities of this type are,
\be \label{possible10fermterms}
f_0(r)H^{\mu\nu\lambda}T_{\mu\nu\lambda} + f_1(r)\Phi^IH^{\mu\nu\lambda}T_{\mu\nu\lambda}^I + \ldots + f_3(r)\Phi^{I} \Phi^{J} \Phi^{K} H^{\mu\nu\lambda}T_{\mu\nu\lambda}^{IJK},
\ee
where $T_{\mu\nu\lambda}^{I_1I_2\ldots I_k}$ is a product of ten fermions with the given antisymmetrized Lorentz indices and traceless symmetrized $Spin(5)$ indices. The strategy we follow will be to show that a specific variation of this top form can only be cancelled by variations of terms coming from the four derivative action and, in particular, that this variation cannot vanish on its own. This is an extension of the top form reasoning of~\cite{Paban:1998ea}. 

The lowest order supersymmetry variation acting on a scalar in these terms results in a term with $1$ $H$-field and $11$ fermions. Such a variation must either vanish on its own, or cancel against the variation of another term in the action; schematically:
\be
\Psi\gamma^I\frac{\partial}{\partial \Phi^I}\left(f(\Phi)H\Psi^{10}\right) = \text{sources}. \label{twelvefermconstraint}
\ee
What are the possible sources that contain a single $H$-field and $11$ fermions? Since there is no $12$ fermion term in the action, no other free supersymmetry variation of the terms in~$\mathcal{L}_3$ will produce a coupling with an $H$ and $11$ fermions. 

What about corrections to the supersymmetry variations acting on lower order terms? Such corrections can be induced from either the four or six derivative terms in the action. First consider corrections induced by the four derivative terms. These corrections 
contain at most $6$ fermions. When acting on the two derivative action, they only mix with the variation of terms of order $4$. 
However, they could produce a source for~\C{twelvefermconstraint}\ when acting on the terms of order $4$.

What about the corrections to the supersymmetry transformations induced by $\mathcal{L}_3$?
These corrections contain at most $10$ fermions so there could be an $11$ fermion variation generated by acting with these corrections acting on the fermion kinetic term. However, because this correction contains $10$ fermions, it cannot contain an $H$ since that would increase its order. It also cannot contain a ``naked'' $B$-field because of gauge invariance. 
Therefore, the only possible source terms come from the corrections to the supersymmetry variations induced by $\mathcal{L}_2$ acting on the terms in $\mathcal{L}_2$. Fortunately, those are the terms we just analyzed in section~\ref{determiningeight}.

We have not yet ruled out a solution to the supersymmetry constraint without sources. Let us do that.
Consider the solution to the homogeneous constraint, i.e., the solution to~\C{twelvefermconstraint} with the source terms zero. Acting on the left with $\gamma^J\frac{\partial}{\partial\Psi}\frac{\partial}{\partial\Phi^J}$ leads to:
\be\label{sixharmonic}
\Delta\left(f(\Phi)H\Psi^{10}\right) = 0 \quad \Rightarrow \quad f_k''(r) + \frac{4+2k}{r}f_k'(r) = 0,
\ee
where $f_k(r)$ are the functions appearing in~\C{possible10fermterms}. Conformal invariance requires that $f_k$ be homogeneous of degree $-11 - k$, which is not a solution to~\C{sixharmonic}\ for $k=0,1,2,3$. Therefore, there are no homogeneous solutions to the supersymmetry constraint on the six derivative top form -- this term must be entirely determined by the four derivative source. Note that in the language of the previous section, such a source term would be proportional to $c_{(6)}^2$. This is because the supersymmetry corrections induced by the four derivative action will come with a factor of $c_{(6)}$. When one of these corrections acts on a four derivative term giving a source, the result will be proportional to $c_{(6)}^2$.

\section{Extension to Rank $N$}
In this section, we will extend our preceeding results for the four derivative terms in the effective action to the case of an arbitrary rank $N$ gauge group broken to its diagonal subgroup $U(1)^N$ for the case of Yang-Mills, or the analogous statement for the $(2,0)$ theory. Our analysis extends the work of~\cite{Sethi:1999qv}. For the moment, let us specialize to $4$-dimensional $\mathcal{N}=4$ Yang-Mills; we will see that with only minor adjustments, the same methods will apply to $(2,0)_N$ as well.

On the Coulomb branch, maximally broken $\mathcal{N}=4$ with gauge group $G$ of rank $N$ contains $N$ massless abelian supermultiplets containing, 
\be
A_A^{\mu}, \, \ \psi_A, \, \ \phi^i_A,
\ee
where $A = 1,\,\ 2,\ldots N$ labels the Cartan of $G$. Since these are abelian multiplets, we can parametrize their supersymmetry transformations as follows, 
\begin{align}
\delta_{\epsilon}\phi^i_A = & i \epsilon \tau^i\psi_A + \epsilon N^i_{AB}\psi_B + \cc, \nonumber \\
\delta_{\epsilon}A^{\mu}_A = & i \overline{\epsilon}\sigma^{\mu}\psi_A + \overline{\epsilon} K^{\mu}_{AB} \psi_B +\cc, \\
\delta_{\epsilon} \psi_A = & F^{\mu\nu}_A\sigma_{\mu\nu} \epsilon + 2\tau^i(\sigma_\mu\partial^\mu \phi^i_A)\overline{\epsilon} + M_A\epsilon + L_A\overline{\epsilon}. \nonumber
\end{align}
The corrections
$K$, $L$, $M$ and $N$ are graded by derivative order as before. Conformal symmetry and maximal supersymmetry ensure that the metric on the moduli space is flat in both 4 and 6 dimensions. Again the first corrections to the supersymmetry transformations occur for $K$, $L$, $M$ and $N$ of order 2, 3, 3 and 2, respectively. This implies a maximum number of fermions in $K$, $L$, $M$ and $N$ of 4, 6, 6 and 4. As before, we focus on the eight fermion terms in the four derivative action whose variation must vanish. This leads to following constraint:
\be
\left(i\epsilon\tau^i\psi_A+i\overline{\epsilon}\tau^i\overline{\psi}_A\right)\frac{\partial}{\partial\phi_A^i}\left(f^{(8)}(\phi)\psi^8\right) = 0.
\ee
Since $\epsilon$ is arbitrary, this implies  two separate equations:
\begin{align} \label{susyconstraints}
\tau^i_{ab}\psi^b_A\frac{\partial}{\partial\phi_A^i}\left(f^{(8)}(\phi)\psi^8\right) = & 0, \nonumber \\
(\tau^i)^{ab}\psi_{bA}\frac{\partial}{\partial\phi_A^i}\left(f^{(8)}(\phi)\psi^8\right) = & 0.
\end{align}
It will be convenient in the following discussion to change notation so that $f^{(8)}$ represents not just the moduli dependence, but also the 8 fermions in this term. From the reality condition on the spinors of the $(2,0)$ theory, there would be only one analogous equation in that case.

Unlike the rank 1 case, $f^{(8)}$ now has a much more complicated dependence on the moduli and fermions since there are $N$ types of each field. Let us choose a particular direction in the Cartan algebra,  say $A = 1$, and grade the 8 fermion terms according to how many $\psi_1$ fermions they contain:
\be
f^{(8)}(\phi) = \sum_{i=0}^8f^{(8)}_i(\phi).
\ee
Each $f^{(8)}_i$ contains $i$ fermions associated with the $A=1$ Cartan direction. The first of our invariance equations then implies,
\be
\tau^i_{ab}\psi_1^b\frac{\partial}{\partial\phi_1^i}\left(f^{(8)}_7(\phi)\right) + \sum_{A\neq1}\tau^i_{ab}\psi_A^b\frac{\partial}{\partial\phi_A^i}\left(f^{(8)}_8(\phi)\right) =  0.
\ee
Multiplying this equation by $\psi_1^a$ and summing over $a$ yields
\be
\sum_{A\neq1}\psi_1^a\tau^i_{ab}\psi_A^b\frac{\partial}{\partial\phi_A^i}\left(f^{(8)}_8(\phi)\right) = 0,
\ee
where we have used
\be
\psi_1^a\tau^i_{ab}\psi_1^b \equiv \psi_1^{\alpha a}\tau^i_{ab}\varepsilon_{\alpha\beta}\psi_1^{\beta b} = 0,
\ee
since both $\tau^i$ and $\varepsilon$ are antisymmetric in their indices. In six dimensions, an analogous relation holds.

Now since $f^{(8)}_8$ only contains $\psi_1$, the contribution from each $A\neq1$ must separately vanish:
\be \label{rankNconstraint}
\tau^i_{ab}\psi_1^b\frac{\partial}{\partial\phi_A^i}\left(f^{(8)}_8(\phi)\right) = 0,
\ee
where $A=1$ is also included since it is implied by the vanishing of the term with 9 $\psi_1$. Continuing as in~\cite{Sethi:1999qv}, we apply the operator $(\tau^j)^{ca}\phi_A^j\left(\frac{\partial}{\partial\psi_{1c}}\right)$ to the above equation while simultaneously applying the conjugate operator to the conjugate equation. Summing these two and summing over $a$ gives,
\be
\left(8\phi^i\frac{\partial}{\partial\phi^i_A} + \phi^i_A\frac{\partial}{\partial\phi^j_A}\left((\tau^{ij})^c_a\frac{\partial}{\partial\psi^c_{1}}\psi_1^a + (\tau^{ij})_c^a\frac{\partial}{\partial\overline{\psi}_{c1}}\overline{\psi}_{a1}\right)\right)f^{(8)}_8 = 0,
\ee
where we used 
\be\left(\frac{\partial}{\partial\psi^c_{1}}\psi_1^c + \frac{\partial}{\partial\overline{\psi}_{c1}}\overline{\psi}_{c1} \right)f^{(8)}_8=8f^{(8)}_8.\ee

Since $f^{(8)}_8$ is $Spin(6)$ invariant, a $Spin(6)$ rotation on the bosons must be compensated by the corresponding rotation acting on the fermions:
\be \label{rotoperators}
\sum_A\left(\phi^i_A\frac{\partial}{\partial\phi^j_A} - \phi^j_A\frac{\partial}{\partial\phi^i_A}\right)f^{(8)}_8 = -\hlf\left(\psi_1 \tau^{ij} \frac{\partial}{\partial\psi_1} + \overline{\psi}_1\tau^{ij}\frac{\partial}{\partial\overline{\psi}_1} \right)f^{(8)}_8.
\ee
After summing on $A$, we therefore find
\be
\left(8r\frac{\partial}{\partial r} - \half\sum_{i<j}\left(\tau^{ij}\frac{\partial}{\partial\psi_1}\psi_1 + \tau^{ij}\frac{\partial}{\partial\overline{\psi}_1}\overline{\psi}_1 \right)^2\right)f^{(8)}_8=0,
\ee
where $r\equiv\sum_{i,A}(\phi^i_A)^2$. This can be rewritten as
\be
\left(8r\frac{\partial}{\partial r} +2\rho_1(C)\right)f^{(8)}_8 = 0,
\ee
where $\rho_1(C)$ is the value of the $Spin(6)$ Casimir in the product of 8 $\psi_1$ appearing in $f^{(8)}_8$.

The most general $f^{(8)}_8$ has a product of 8 $\psi_1$ fermions contracted with scalars $\phi^i_A$ to make a $Spin(6)$ scalar. 
 Any $Spin(6)$ invariant function of scalars can multiply this structure. However, conformal invariance requires that $f^{(8)}_8$ be a homogeneous function of the scalars of degree $-8$. This implies that $\rho_1(C) = 32$. In six dimensions, the homogeneity would be of degree $-7$, giving $\rho_1(C) = 28$. Armed with this constraint from conformal invariance, we can deduce  the highest weight of the $Spin(6)$ representations appearing in $f^{(8)}_8$. 

For the rank $3$ group $Spin(6)$, we choose an orthonormal basis on the weight space of $\{w_1,w_2,w_3\}$ with positive roots $w_i \pm w_j$, $i<j$. The sum of positive roots is
\be
2\delta = 4w_1+2w_2.
\ee
For a representation with highest weight $\lambda$, the value of the Casimir in that representation is, 
\be
\rho(C) = \langle\lambda + 2\delta, \lambda\rangle.
\ee
Let the highest weight of the representations appearing in $f^{(8)}_8$ be expressed as 
\be aw_1 + bw_2 + cw_3 \ee 
for integers $a,b,c$ satisfying $a\geq b \geq |c|$. Then,
\bea
32 &=& \langle(4+a)w_1 + (2+b)w_2 + cw_3, aw_1 + bw_2 + cw_3\rangle, \cr &=& a(a+4) + b(b+2) + c^2,
\eea
whose only solution is $a=4$, $b=c=0$. This is the traceless, totally symmetric 4 index tensor representation, which is precisely the representation which appeared previously for a rank $1$ gauge group. The preceding analysis can be applied straightforwardly to six dimensions, which yields agreement with the rank 1 case as well by restricting to the symmetrized, traceless $4$ index tensor representation.

It will be useful for us to arrive at this same conclusion in another way.\footnote{We would like to thank M.~Stern for clarification on this and subsequent points.} 
Choose the $Spin(6)$ generator $\tau^{12}$ to be an element of the Cartan subalgebra, and also choose coordinates in the weight space such that the basis vector $w_1$ is dual to $\tau^{12}$. We can then decompose the fermions into eigenvectors of the fermionic 1 - 2 rotation generator appearing in~\C{rotoperators},
\begin{align}
\psi^a_A &=  \psi^{a+}_A + \psi^{a-}_A, \nonumber \\
\overline{\psi}_{aA} &=  \overline{\psi}^{+}_{aA} + \overline{\psi}^{-}_{aA},
\end{align}
where the ``$+$'' fermions have eigenvalue $i/2$ and the ``$-$'' fermions have eigenvalue $-i/2$ when acted on by the 1 - 2 component of the rotation operator appearing in~\C{rotoperators}. Note that these eigenvalues are a consequence of $\tau^{12}$ squaring to $-I$.

Additionally, we define momenta conjugate to the bosons, $\phi^i_A$:
\be
\com{\phi^i_A}{\Pi^j_A} = i\delta^{ij}\delta_{AB}.
\ee
Under the 1 - 2 generator of $Spin(6)$ rotations appearing in~\C{rotoperators}, the momenta $\Pi^i_A$ for $i\neq 1,2$ have eigenvalue zero. The two momenta with $i=1,2$ can be written in terms of eigenvectors as follows,
\begin{align}
\Pi^1_A &= \frac{1}{2}\left(\partial_{z_A} + \partial_{\bar{z}_A}\right), \nonumber \\
\Pi^2_A &= -\frac{i}{2}\left(\partial_{z_A} - \partial_{\bar{z}_A}\right),
\end{align}
where $\partial_{z_A}$ and $\partial_{\bar{z}_A}$ have eigenvalues $-i$ and $i$, respectively. The supersymmetry constraints~\C{susyconstraints} can be written
\begin{align}
Q_af^{(8)} = \overline{Q}_a & f^{(8)} = 0, \\ \nonumber
Q_a = i\tau^{i}\psi_A\Pi^i_A, & \qquad \overline{Q}_a = i\tau^i\overline{\psi_A}\Pi^i_A,
\end{align}
where the supercharges $Q_a$, $\overline{Q}_a$ each decompose into four pieces which act differently on the $w_1$ component of the weights of the fermionic and bosonic operators composing $f^{(8)}$. 

The four pieces, which we will label as $Q_a^{(+,i)}$ and $Q_a^{(-,i)}$, with $i=1,2$ (with similar expressions for $\overline{Q}_a$) act in the following way: $Q_a^{(+,1)}$ and $Q_a^{(+,2)}$ raise the total $w_1$ component of the weight by $1/2$. The first raises the bosonic contribution by 1 and lowers the fermionic contribution by $1/2$ while the second raises the fermionic contribution by $1/2$ and does not change the bosonic contribution; $Q_a^{(-,1)}$ and $Q_a^{(-,2)}$ lower the total $w_1$ component of the weight by $1/2$, with the first lowering the bosonic contribution by 1 and raising the fermionic contribution by $1/2$, while the second lowers the fermionic contribution by $1/2$ and does not change the bosonic contribution.

Since $f^{(8)}_8$ is a $Spin(6)$ scalar, it is both a highest and lowest weight state and it must be annihilated by each of the $Q$ and $\overline{Q}$ separately. In particular, it must be annihilated by the piece in $Q_a^{(-,1)}$ that contains only the first Cartan element which can be written,
\be \label{Q-1operator}
Q_a^{(-,1)} \supset dz_{1\Omega}\partial_{z_1},
\ee
where $dz_{1\Omega}$ is a certain linear combination of $\psi^{a+}_1$ and $\Omega = 1,2,3,4$. That this operator annihilates $f^{(8)}_8$ implies that it is either anti-holomorphic in $z_1$ or a product of all four $dz_{1\Omega}$. However, if $f^{(8)}_8$ were anti-holomorphic in $z_1$, it must necessarily be constant since it is bounded almost everywhere, and the only physical singularities in the moduli space are real codimension $6$. A constant solution is unphysical since $f^{(8)}_8$ must vanish as we approach infinity along any non-singular direction. Therefore, 
\be 
f^{(8)}_8 \propto \prod_{\Omega = 1}^4 dz_{1\Omega}.
\ee
Additionally, consideration of the analogous constraint from $\overline{Q}_a^{(-,1)}$ gives
\be
f^{(8)}_8 \propto \prod_{\Omega = 1}^4\prod_{\Lambda = 1}^4 dz_{1\Omega}d\overline{z}_{1\Lambda},
\ee
where $d\overline{z}_{1\Lambda}$ is the object corresponding to $dz_{1\Lambda}$ but composed of $\overline{\psi}_1^{a+}$. This product of fermions has highest weight of $4w_1$ as required by conformal invariance.

Furthermore, since $f^{(8)}_8$ is proportional to all eight ``$+$'' type fermions, the entire eight fermion term $f^{(8)}$ vanishes if $f^{(8)}_8$ does. This is because, with $f^{(8)}_8=0$, each component of $f^{(8)}_7$ must be annihilated by $dz_{1\Omega}\partial_{z_1}$; however, this implies that $f^{(8)}_7 \propto f^{(8)}_8 =0$. This reasoning iterates for each $f^{(8)}_i$. All of $f^{(8)}$ is then determined via supersymmetry from the known function $f^{(8)}_8$ whose fermion-dependence is identical to the rank $1$ situation.

All of the preceding reasoning applies to the $6$-dimensional rank $N$ theory with changes only in notation but not in content. In the rank 1 analysis, we were able to completely determine the eight fermion term up to a single constant. Here so far, we have determined the rank $N$ eight fermion couplings in terms of a function $f^{(8)}_8$ of the moduli. We would now like to argue that it is again true that we can fix the eight fermion terms in 4 and 6 dimensions up to a constant and further that these constants are related.

\subsection{Solving for $f^{(8)}_8$ in 4 and 6 dimensions}
We can do more than just determine the representation appearing in $f^{(8)}_8$; we can, in fact, determine $f^{(8)}_8$ up to a constant as in the rank $1$ case. We will use the analogue of the harmonicity constraint from the rank $1$ analysis, which we deduce by acting on~\C{rankNconstraint} with the operator $\tau^j\frac{\partial}{\partial \psi_1}\frac{\partial}{\partial \phi^j_A}$, with no sum on $A$. This gives a collection of equations,
\be \label{pluriharm}
\sum_{i=1}^6\left(\frac{\partial}{\partial \phi^i_A}\right)^2f^{(8)}_8=0,
\ee
whose solutions are functions termed pluri-harmonic in~\cite{Sethi:1999qv}. Let us focus on the $A=1$ case. In gauge theory, we know that the eight fermion term is only allowed to diverge when a non-abelian symmetry appears. The same singularity structure is true for the $(2,0)$ theory. 

For an $A_N$ gauge group and $A=1$, these loci correspond to $\phi^i_1= \phi^i_B$ for some $B\neq 1$. For a general gauge group, we parametrize these loci as follows~\cite{Buchbinder:1998qd}: introduce the Weyl basis $\{H_A, E_{+\hat{\alpha}},E_{-\hat{\alpha}}\}$. The $H_A$, $A=1,\ldots N$ are the generators of the Cartan and $\pm \hat{\alpha}$ are the positive and negative roots. For a rank $N$ gauge group completely broken to $U(1)^N$, the massless scalar fields are valued in the Cartan, 
\be
\phi^i = \sum_A\phi^i_AH_A,
\ee
where this defines $\phi^i_A$ used previously. A non-abelian gauge symmetry generated by $E_{\hat{\alpha}}$ is restored if $\com{\phi^i}{E_{\hat{\alpha}}}=0$. If we write
\be
\hat{\alpha} = \sum_A\alpha_Aw_A,
\ee
where $w_A$ are the basis vectors in the root space dual to the $H_A$ (not to be confused with the basis of the $Spin(6)$ $R$-symmetry root space introduced earlier), then the vanishing of the commutator $\com{\phi^i}{E_{\hat{\alpha}}}$ corresponds to
\be
\phi^i_{\hat{\alpha}}   \equiv  \sum_A\alpha_A\phi^i_A=0.
\ee

For an $A_N$ gauge group, this reduces to our previous assertion since each root $\hat{\alpha}$ is given by the difference of two basis vectors in (the traceless subspace of) the root space $\mathbb{R}^{N+1}$. The general solution to the pluri-harmonicity constraint with singularities only at loci of enhanced symmetry, and which also vanishes as $\phi_1\to\infty$ in almost all directions, is given by:
\be
f^{(8)}_8 = \sum_{\hat{\alpha} \in \Lambda_1}\frac{c_{\hat{\alpha}}\phi^i_{\hat{\alpha}}\phi^j_{\hat{\alpha}}\phi^k_{\hat{\alpha}}\phi^l_{\hat{\alpha}}}{|\phi_{\hat{\alpha}}|^{12}}T^{ijkl}.
\ee
The sum here is over roots in the subspace $\Lambda_1$ of root space defined such that for each $\hat{\alpha} \in \Lambda_1$, $\alpha_1> 0$, i.e., $\Lambda_1$ consists of those positive roots with a nonzero $w_1$ component. We restrict to this subspace in order to ensure the vanishing of this eight fermion term as $|\phi_1|\to\infty$.
 
In fact, for a simply-laced classical gauge group, the action of the Weyl group ensures that $c_{\hat{\alpha}} = c_{(4)}$ for any $\hat{\alpha}\in\Lambda_1$, where $c_{(4)}$ depends on the group $G$. Interestingly, it is known that for $4$-dimensional $\mathcal{N} = 4$ Yang-Mills,  each $c_{\hat{\alpha}}$ is determined by a single constant $c_{(4)}$ for any gauge group~\cite{Buchbinder:1998qd}.

The same arguments in six dimensions lead to:
\be
f^{(8)}_8 = c_{(6)}\sum_{\hat{\alpha} \in \Lambda_1}\frac{\Phi^I_{\hat{\alpha}}\Phi^J_{\hat{\alpha}}\Phi^K_{\hat{\alpha}}\Phi^L_{\hat{\alpha}}}{|\Phi_{\hat{\alpha}}|^{11}}T^{IJKL}.
\ee
Here we have made the replacement $c_{\hat{\alpha}} =c_{(6)}$. 
If this were not true, we will find disagreement with the $4$-dimensional result on torus compactification. 

\subsection{Connecting $c_{(4)}$ and $c_{(6)}$}

We will proceed as in the rank 1 analysis to show that $c_{(4)}$ and $c_{(6)}$ are simply related by studying the toroidally compactified $(2,0)_N$ theory where the torus $T^2$ again has volume $V_T$. On the torus, we have the same supersymmetry constraint on the eight fermion term as in 4 dimensions, since the free particle supersymmetry variations cannot detect the existence of a compact direction in the moduli space:
\begin{align}
\tau^i_{ab}\psi^b_A\frac{\partial}{\partial\phi_A^i}\left(f^{(8)}(\phi)\right) = & 0, \nonumber \\
(\tau^i)^{ab}\psi_{bA}\frac{\partial}{\partial\phi_A^i}\left(f^{(8)}(\phi)\right) = & 0.
\end{align}
The sum over $i$ includes both the non-compact directions $i=1,\ldots5$ and the compact direction $i=6$. We again grade the eight fermion term according to the number of preferred fermions $\psi_1$, and as before, we arrive at  constraints on $f^{(8)}_8$:
\be \label{rankNtorusconstraint}
\tau^i_{ab}\psi_1^b\frac{\partial}{\partial\phi_A^i}\left(f^{(8)}_8(\phi)\right) = 0.
\ee
These constraints imply pluri-harmonicity of $f^{(8)}_8$ on the moduli space. 

As can be checked, the arguments which led to the assertion that $f^{(8)}_8$ has a highest weight of $4w_1$ are also independent of the shape of the moduli space. Therefore $f^{(8)}_8$, as a representation of $Spin(6)$, is in the traceless symmetric four component tensor representation. However, under the manifest $Spin(5)$ subgroup, this decomposes into the sum of traceless, symmetric, $k$ component tensor representations for $0\leq k\leq4$. The pluri-harmonic constraint acting on these structures will produce a sequence of Bessel equations as we saw in the rank $1$ analysis.

At this point, we will restrict to classical simply-laced groups. The exceptional groups can be handled in a similar way, but there are some potential complications that we will avoid here. Let us Fourier expand in each of the periodic scalars $\phi^6_A$,
\be \label{fourierexpansion}
f^{(8)}_8 = \sum_{\vec{n}}f^{(8)}_{8,\vec{n}}e^{i\vec{n}\cdot\vec{\phi}^6\sqrt{V_T}},
\ee
where $\vec{n}\cdot\vec{\phi}^6 = \sum_A n_A\phi^6_A$ and $n_A\in\mathbb{Z}$. Each $f^{(8)}_{8,\vec{n}}$ is itself a sum of contributions from each of the $k$ scalar structures. If we focus on the $\vec{n}=0$ mode and the four scalar structure then the solution with appropriate singularities and asymptotic behavior is
\be
f^{(8)}_8 = c_{(6)}V_T^{1/2}\sum_{\alpha \in \Lambda_+}\frac{\phi^I_{\alpha}\phi^J_{\alpha}\phi^K_{\alpha}\phi^L_{\alpha}}{|\phi_{\alpha}|^{11}}T^{IJKL}.
\ee
The coefficient $c_{(6)}$ in this expression is the same as we would find in the case of six non-compact dimensions; this can be seen from taking the decompactification limit. Furthermore, we again find that our solution for the non-zero Fourier modes implies that the zero mode contributions from structures with fewer than $4$ scalars must vanish, so we ignore them.

Next focus on the non-zero modes and consider the weaker constraint derived from~\C{pluriharm}\ by summing over the Cartan label $A$:
\be
\sum_A\Delta_Af^{(8)}_8 = \sum_A\left(\Delta_A' + \left(\frac{\partial}{\partial\phi^6_A}\right)^2\right)f^{(8)}_8 =0.
\ee
In the first equality we have defined the operator $\Delta_A'$ to be the Laplacian over the non-compact part of the moduli space associated with the $A^{\text{th}}$ Cartan element. Substituting the expansion of $f^{(8)}_8$ found in~\C{fourierexpansion} gives:
\be
\sum_{\vec{n}}\left(\sum_A\Delta_A'-|\vec{n}|^2V_T\right)f^{(8)}_{8,\vec{n}}e^{i\vec{n}\cdot\vec{\phi}^6\sqrt{V_T}}=0.
\ee
Linear independence of the exponentials implies:
\be
\left(\sum_A\Delta_A'-|\vec{n}|^2V_T\right)f^{(8)}_{8,\vec{n}}=0.
\ee


For the pluri-harmonic solutions in non-compact space, we found that $f^{(8)}_8$ was given by a sum of functions each depending only on a single $\phi_{\hat{\alpha}}$ associated with a root $\hat{\alpha}\in\Lambda_1$. This must again be the case for the non-compact scalars, $\phi^I$, with manifest $Spin(5)$ symmetry in order to agree with the $4$-dimensional limit.  We will see that supersymmetry requires the same structure for the compact scalars. 



Therefore, we make the expansion
\be
f^{(8)}_{8,\vec{n}}(\phi_A^I) = \sum_{\hat{\alpha}\in\Lambda_1}f^{(8),\hat{\alpha}}_{8,\vec{n}}(\phi_{\hat{\alpha}}^I).
\ee
Note we are not using the stronger claim that even the dependence on $\phi^6_A$ is organized this way. We will nevertheless see that the pluri-harmonic constraint demands that this is the case. Making a change of basis,\footnote{For the $D_N$ case, we should restrict the sum in~\C{changeofbasis}\ over ${\hat{\alpha}}$ to roots in $\Lambda_1$ of the form $w_1+w_j$ for $j>1$. The Weyl group contains an element relating  $w_1+w_j$ to $w_1-w_j$ which can be used to generate the remaining ${\hat{\alpha}} \in \Lambda_1$. This restriction removes any worries about linear dependence in making the change of basis. }
\be\label{changeofbasis}
\sum_A\Delta_A' = \sum_{\hat{\alpha}}|\hat{\alpha}|^2\Delta_{\hat{\alpha}}',
\ee
we arrive at the weaker harmonic constraint
\be
\left(|\hat{\alpha}|^2\Delta_{\hat{\alpha}}'-|\vec{n}|^2V_T\right)f^{(8),\hat{\alpha}}_{8,\vec{n}}=0.
\ee

Alternatively, we could have considered the full pluri-harmonicity requirement by not summing over $A$. With the same ansatz, this would lead to:
\be
\left((\alpha_A)^2\Delta_{\hat{\alpha}}'-(n_A)^2V_T\right)f^{(8),\hat{\alpha}}_{8,\vec{n}}=0.
\ee
For both constraints to be simultaneously satisfied, we must insist that $f^{(8),\hat{\alpha}}_{8,\vec{n}}$ vanishes unless $\vec{n}\propto\hat{\alpha}$. This allows us to rewrite the Fourier expansion of $f^{(8)}_8$ as follows:
\be
f^{(8)}_8 = \sum_{\hat{\alpha}\in\Lambda_1}\sum_{n_{\hat{\alpha}}}f^{(8),\hat{\alpha}}_{8,n_{\hat{\alpha}}}e^{in_{\hat{\alpha}}\phi_{\hat{\alpha}}^6\sqrt{V_T}/|\hat{\alpha}|}.
\ee
For the simply-laced Lie algebras we are considering, $|\hat{\alpha}| = \sqrt{2}$. We will make this replacement later. 


Now focus on the eight fermion term for just one $\hat{\alpha}\in\Lambda_1$ since the Weyl group will determine the rest. Separate $f^{(8),\hat{\alpha}}_{8,n_{\hat{\alpha}}}$ into its contributions from the $k$ scalar structures,
\be
f^{(8),\hat{\alpha}}_{8,n_{\hat{\alpha}}} = \sum_kf^{\hat{\alpha}}_{k,n_{\hat{\alpha}}}(r_{\hat{\alpha}})\phi_{\hat{\alpha}}^{I_1}\ldots\phi_{\hat{\alpha}}^{I_k}T^{I_1\ldots I_k},
\ee
where we define the $Spin(5)$ invariant combination $r_{\hat{\alpha}} = \sqrt{\phi^I_{\hat{\alpha}}\phi^I_{\hat{\alpha}}}$. The Laplace equation now reads:
\be
\left(\frac{d^2}{dr_{\hat{\alpha}}^2} + \frac{2k + 4}{r_{\hat{\alpha}}}\frac{d}{dr_{\hat{\alpha}}} - \frac{n_{\hat{\alpha}}^2V_T}{2}\right)f^{\hat{\alpha}}_{k,n_{\hat{\alpha}}} = 0.
\ee
This is exactly the Bessel equation that appeared in~\C{rank1bessel}. In fact, from this point forward the analysis proceeds exactly as in section~\ref{rank1torus}. In particular, we can use~\C{rankNtorusconstraint}\ to arrive at the analogue of~\C{orderconstraint}, which relates all of the $k$ scalar structures at leading order in $1/r_{\hat{\alpha}}$. This relation is,
\be
\left(\frac{1}{r_{\hat{\alpha}}}\phi^I_{\hat{\alpha}}\tau^I - i \mathrm{sign}(n_{\hat{\alpha}})\tau^6\right)\psi_1h_1 =\left(\frac{1}{r_{\hat{\alpha}}}\phi^I_{\hat{\alpha}}\tau^I - i \mathrm{sign}(n_{\hat{\alpha}})\tau^6\right)\overline{\psi}_1h_1=0,
\ee
where $h_1$ consists of those $(\psi_1)^8$ terms with leading $1/r_{\hat{\alpha}}$ behavior, as before. Continuing in this way, we see that the $k$ scalar structures at level $n_{\hat{\alpha}}$ are related and completely determined up to a coefficient $c_{n_{\hat{\alpha}}}$. Again, comparison with the $4$-dimensional coefficient via Poisson resummation gives:
\be
c_{(4)} \propto c_{n_{\hat{\alpha}}} \propto c_{(6)}.
\ee

\section{The Conformal Anomaly of $(2,0)_N$}
As we have shown, the eight fermion terms of the $(2,0)_N$ theory with the analogue of gauge group $G$ broken to its maximal torus are determined by their dimensional reduction to $\mathcal{N}=4$ supersymmetric Yang-Mills theory. As we discussed in section~\ref{eightfortwozero}, we expect in both $4$ and $6$ dimensions that the eight fermion terms determine the rest of the four derivative couplings by supersymmetry. Under this assumption, we will consider the whole four derivative action to be multiplied by a coefficient ($c_{(4)}$ or $c_{(6)}$ in our prior discussion). The value of this coefficient can be determined by comparison with a known loop result.

Consider the four derivative action in four dimensions. So far, our results have not required choosing any special values for the moduli. Let us now specialize to a particular corner of the moduli space where the VEV of the scalar associated with our preferred Cartan direction, $\phi^i_1$, is much larger than any other VEV: $v_1 \gg v_A$, $\, A\neq 1$. Additionally, choose the other VEVs to be approximately equal. The effective action is then well approximated by that of the gauge group $G$ broken to $H\times U(1)$, where the $U(1)$ factor is associated with the preferred Cartan direction. In a brane picture, we cluster $N-1$ branes close to each other and take a single brane far away. This kind of limit was considered in~\cite{Dine:1999jq}. In this limit, we can separate terms in the effective action
\be
\mathcal{L}_{\text{eff}} \sim \mathcal{L}_{U(1)} + \mathcal{L}_{H} + \mathcal{L}_{\text{mixed}},
\ee
where the terms in $\mathcal{L}_{\text{mixed}}$ are composed of products of approximately $H$-invariant operators and $U(1)$ fields. A discussion of mixed operators of this type in relation to the $a$-theorem can be found in~\cite{Nakayama:2011wq}.

We will also change our basis for the the moduli to include the dilaton $\tau$, which is the Nambu-Goldstone boson associated with the spontaneous breaking of conformal symmetry,
\be
e^{-2\tau} = \frac{1}{v^2}\left((\phi^i_1)^2 + (\phi^i_2)^2 +\ldots + (\phi^i_N)^2\right) \approx \frac{1}{v_1^2}\left(\phi^i_1\right)^2,
\ee
where $v^2 = \sum_Av_A^2 \approx v_1^2$. In this basis, we find that  the terms appearing in $\mathcal{L}_{\text{mixed}}$ at order $4$ are small in the limit we are taking since they are suppressed by powers of the large VEV $v_1$ relative to other terms in the action. A typical example of such a term is
\be
\mathcal{L}_{\text{mixed}} \supset (\partial \tau)^2\mathcal{O}_{\Delta}\left(\frac{e^{\tau}}{v_1}\right)^{\Delta-2}
\ee
where $\Delta$ is the dimension of the $H$-invariant operator $\mathcal{O}_{\Delta}$. To appear in the four derivative action, $\mathcal{O}_{\Delta}$ must contain two derivatives and some fields, giving $\Delta>2$, so this example of a mixed coupling is suppressed. A four derivative term in $\mathcal{L}_{U(1)}$, on the other hand, has the following form
\be
\mathcal{L}_{U(1)} \supset \frac{(\partial \phi_1)^4}{(\phi_1)^4} \approx (\partial\tau)^4
\ee
which is not suppressed by $v_1$.

A more general statement is that the terms in $\mathcal{L}_{\text{mixed}}$ necessarily contain more powers of $\phi_1$ in the denominator than in the numerator, and therefore will be suppressed by powers of $v_1$ when we change to the dilaton basis. This fact allows us to ignore these terms in the following analysis. Anomaly matching~\cite{Schwimmer:2010za, Komargodski:2011vj}\ tells us that the coefficient $c_{(4)}$ is proportional to the difference in conformal anomalies of the $G$ and $H\times U(1)$ theories,
\be
c_{(4)} \propto \Big(a_4(G) - a_4(H\times U(1))\Big),
\ee
where $a_4(G)$ is the conformal anomaly of $\mathcal{N} =4$ SYM with gauge group $G$. This in turn tells us that in six dimensions, the four derivative action for the $U(1)$ factor is also proportional to this difference,
\be
\mathcal{L}_{U(1)} \sim \Big(a_4(G) - a_4(H\times U(1))\Big)\Big((\partial\tau)^4V_1e^{-\tau} + \ldots \Big),
\ee
where $V_1$ is the large VEV of the six dimensional $\Phi_1$ field, and included in the omitted terms are the eight fermion terms we have explicitly computed. 

We have previously shown that for a rank $1$ gauge group, the $10$ fermion term in six dimensions (i.e., the top form in $\mathcal{L}_3$)  is completely determined by the four derivative action. Under the assumption that supersymmetry connects all  terms in the rank  $1$ six derivative action (which is known to be true for $0+1$ maximal supersymmetric Yang-Mills~\cite{Sethi:2004np}), this determines the six derivative action completely in terms of the four derivative action. We would next like to argue that for the choice of VEVs we are considering, the $U(1)$ factor of the four derivative action completely determines the $U(1)$ factor of the six derivative action up to corrections suppressed by the VEV.

As in the rank $1$ analysis, there is no homogeneous solution to the supersymmetry constraint on the top form of the $U(1)$ factor that is compatible with conformal invariance. Therefore, as before, the top form must be completely sourced by the four derivative action. Certainly variations of the four derivative $U(1)$ factor will contribute. What about variations from $\mathcal{L}_H$ or $\mathcal{L}_{\text{mixed}}$? Since $\mathcal{L}_H$ contains no fields associated with the preferred Cartan direction, it is impossible to have a variation of a term in $\mathcal{L}_H$ that will contains only fields in the preferred direction. The same argument applies to $\mathcal{L}_{\text{mixed}}$, which must contain fields in an approximately $H$-invariant combination. Therefore, only the variation of order $4$ terms in $\mathcal{L}_{U(1)}$  appropriately source  the order $6$ $U(1)$ action.
These order $6$ terms must then take the form,
\be\label{sixcoefficient}
\mathcal{L}_{U(1)} \sim \Big(a_4(G) - a_4(H\times U(1))\Big)^2\Big((\partial\tau)^6 + \ldots\Big),
\ee
where the ellipses again include the top forms we studied previously. In Appendix~\ref{computeanomalies}, we have extended the analysis of~\cite{Schwimmer:2010za, Komargodski:2011vj}\ to six dimensions. Using equation~\C{6danomalyaction}, we can read of the difference in the conformal anomaly of the $(2,0)$ theory associated to gauge group $G$ and to gauge group $H\times U(1)$ from~\C{sixcoefficient}\ to find: 
\be
a_6(G) - a_6(H\times U(1)) \propto \Big(a_4(G) - a_4(H\times U(1))\Big)^2.
\ee
There are potential (group independent) overall numerical factors in~\C{sixcoefficient}\ which can be determined by more carefully studying the supersymmetry variations, but we do not need them to determine the leading behavior of the anomaly.  
For $G = SU(N+1)$ and $H=SU(N)$, we find that $a_6(SU(N+1)) - a_6(SU(N)\times U(1)) \propto N^2$. This  implies that the anomaly for an unbroken rank $N$ $(2,0)$ theory has the long sought leading behavior of $N^3$ at large $N$. With more control over the couplings and supersymmetry transformations of the effective action, it should be possible to determine the exact value of the conformal anomaly.

\section*{Acknowledgements}

It is our pleasure to thank M.~B. Green, Z.~Komargodski, D.~Kutasov and particularly M.~Stern for discussions.
T.~M. is supported in part by NSF Grant No.~PHY-0758029 and by a GAANN Fellowship. S.~S. is supported in part by
NSF Grant No.~PHY-0758029 and NSF Grant No.~0529954. 

\appendix
\section{Conformal Anomalies from 6D Effective Actions}\label{computeanomalies}

In order to compute the conformal anomaly from a Coulomb branch effective action, we must extend the work of~\cite{Schwimmer:2010za, Komargodski:2011vj}\ to  six dimensions. 
It is natural to switch to a basis of fields that includes the dilaton, $\tau$, which is the Nambu-Goldstone boson of spontaneously broken conformal symmetry.  We would like to classify all possible terms in the effective action for the dilaton up to and including six derivatives. 

We begin by finding all such terms that are both diffeomorphism and Weyl invariant by introducing a Weyl invariant metric as follows: consider the Weyl variation of the metric and the dilaton:
\be
g_{\mu\nu}\longrightarrow e^{2\sigma}g_{\mu\nu}, \,\ \tau \longrightarrow \tau +\sigma.
\ee
Under the action of diffeomorphisms, the metric transforms as usual while $\tau$ is a scalar. The combination $\hat{g}_{\mu\nu} = e^{-2\tau}g_{\mu\nu}$ will then transform as a standard metric under diffeomorphisms but is also Weyl invariant. The diff$\times$Weyl invariant terms in the effective action can then be written in terms of the diffeomorphism invariants of the metric $\hat{g}$.

Up to two derivatives, we have the couplings
\be
S_2 = \alpha^4 \int d^6x\sqrt{-\hat{g}}\left(\Lambda + \frac{\hat{R}}{20}\right),
\ee
where $\Lambda$ is a cosmological constant, $\hat{R}$ is the Ricci scalar of $\hat{g}$, and $\alpha$ is a dimension $1$ constant analogous to a decay constant in pion physics. Following~\cite{Komargodski:2011vj}, the cosmological constant term is ruled out on the grounds that its presence conflicts with spontaneously broken conformal symmetry. For flat space where $\hat{g}_{\mu\nu} = e^{-2\tau}\eta_{\mu\nu}$, the two derivative term gives (after integrating by parts)
\be
S_2 = \alpha^4 \int d^6x \, e^{-4\tau}(\partial \tau)^2,
\ee
which leads to the equation of motion $\Box\tau = 2(\partial \tau)^2$.

At four derivatives, there are three independent combinations of curvatures:
\be
S_4 = \int d^6x\sqrt{-\hat{g}}\left(\kappa_1\hat{R}^2 +\kappa_2\hat{E}_4 + \kappa_3\hat{W}_{\mu\nu\rho\sigma}^2\right),
\ee
where the constants $\kappa_i$ have dimension $2$, $E_4$ is the $4$-dimensional Euler density, and $W$ is the Weyl tensor. In flat space, the last term gives no contribution since it is conformally invariant, and $\hat{g}$ is conformal to the Minkowski metric. Additionally, $\hat{R}^2$ will only give four derivative couplings proportional to the two derivative equation of motion, since $\hat{R}$ itself is proportional to the leading equation of motion. Thus, only the Euler density term will lead to a genuine four derivative coupling. 
This is to be contrasted with the $4$-dimensional case where there are no diff$\times$Weyl invariant four derivative terms that are not proportional to the leading equation of motion. In six dimensions, we will see that all diff$\times$Weyl invariant six derivative terms are proportional to the leading equation of motion.

At six derivatives, there are $10$ independent curvature combinations, up to total derivatives~\cite{Oliva:2010zd}:
\bea \label{6dcurvaturebasis}
&\hat{L}_1 = \hat{W}^{\mu\nu\rho\sigma}\hat{W}_{\rho\sigma\lambda\delta}\hat{W}^{\lambda\delta}_{\mu\nu}, \qquad & \hat{L}_2 = \hat{W}^{\mu\nu}_{\rho\sigma}\hat{W}^{\rho\delta}_{\nu\lambda}\hat{W}^{\sigma\lambda}_{\mu\delta},  \nonumber \\
& \hat{L}_3= \hat{W}^{\mu\nu\rho\sigma}\hat{W}_{\rho\sigma\nu\lambda}\hat{R}^{\lambda}_{\mu}, \qquad & \hat{L}_4 = \hat{W}^{\mu\nu\rho\sigma}\hat{W}_{\mu\nu\rho\sigma}\hat{R}, \nonumber \\
&\hat{L}_5 =  \hat{W}^{\mu\nu\rho\sigma}\hat{R}_{\mu\rho}\hat{R}_{\nu\sigma}, \qquad & \hat{L}_6 = \hat{R}^{\mu\nu}\hat{R}_{\nu\rho}\hat{R}^{\rho}_{\mu}, \\
&\hat{L}_7 =  \hat{R}^{\mu\nu}\hat{R}_{\mu\nu}\hat{R}, \qquad & \hat{L}_8 = \hat{R}^3,  \nonumber \\
&\hat{L}_9 =  \hat{R}\hat{\Box}\hat{R}, \qquad & \hat{L}_{10} = \hat{W}^{\mu\nu\rho\sigma}\hat{\Box}\hat{W}_{\mu\nu\rho\sigma}. \nonumber
\eea
The six derivative action is then
\be
S_6 = \int d^6x \sqrt{-\hat{g}}\sum_{i=1}^{10}\beta_i\hat{L}_i,
\ee
with $\beta_i$ dimensionless constants. In flat space, all of these terms either vanish or are proportional to the leading equation of motion except $\hat{L}_6$. After a tedious calculation, this term is also seen to give a contribution proportional to the leading equation of motion. Therefore, there are no diff$\times$Weyl invariant contributions to the tree level scattering of six dilatons in six dimensions, which is the same as the case of four dilaton scattering in four dimensions.

We next need to consider potential non-Weyl invariant terms which correspond to anomalous contributions to the effective action. In six dimensions, using the terminology of~\cite{Deser:1993yx}, there is one A-type anomaly given by the six-dimensional Euler density $E_6$ and three B-type anomalies from various combinations of Weyl tensors, which we will label $\mathcal{A}_i$ with $i=1,2,3$. In terms of our basis of curvature couplings~\C{6dcurvaturebasis}, these anomalies are
\bea
E_6 &= &2L_1-8L_2+6L_3+\frac{6}{5}L_4+3L_5+\frac{3}{2}L_6-\frac{27}{20}L_7+\frac{21}{100}L_8, \nonumber \\
\mathcal{A}_1 &= & L_1, \\
\mathcal{A}_2 &= & L_2, \nonumber \\
\mathcal{A}_3 &= & L_{10} + 2L_3 -3L_5-\frac{3}{2}L_6 + \frac{27}{20}L_7 -\frac{21}{100}L_8. \nonumber
\eea
The absence of hats signifies that these curvatures are given by~\C{6dcurvaturebasis}\ but are evaluated using the Weyl variant metric $g$, rather than its Weyl invariant counterpart $\hat{g}$. Interestingly, because of the relation between $E_6$ and $L_6$ above, it is clear why $\hat{L}_6$ could not give a contribution to six $\tau$ scattering.

We threfore seek a six derivative action $S_{anomaly}$ that gives the following Weyl variation:
\be
\delta_{\sigma}S_{anomaly} = \int d^6x \sqrt{-g}\sigma\left(aE_6 + \sum_{i=1}^3b_i\mathcal{A}_i \right).
\ee
Note that we have ignored potential anomalous variations that can be removed via local counter-terms, since these variations do not represent true anomalies.

For the B-type anomalies, finding such an action is easy because $\sqrt{-g}\mathcal{A}_i$ is Weyl invariant, so we can simply replace $\sigma$ with $\tau$ in the variation to give the action. We can also replace $\sigma$ with $\tau$ in the A-type anomaly variation, but since $\sqrt{-g}E_6$ is not Weyl invariant, we need extra Wess-Zumino terms in $S_{anomaly}$ to cancel the variation of $\sqrt{-g}E_6$. The precise action can be obtained through a Wess-Zumino procedure performed in~\cite{Arakelian:1995ye},
\bea
S_{anomaly} &=& \sum_{i=1}^{3}b_i\int d^6x\sqrt{-g}\tau\mathcal{A}_i +  a\int d^6x\sqrt{-g}\Bigg( \tau E_6 -\frac{1}{6!}\Big[ 6(\nabla \tau)^6 -18\Box\tau(\nabla \tau)^4 \nonumber \\
&& +24(\nabla \tau)^2(\nabla_{\mu}\nabla_{\nu}\tau\nabla^{\mu}\nabla^{\nu}\tau - (\Box\tau)^2) -6R(\nabla \tau)^4 \nonumber \\
&& -4G^{\mu\nu}_{\lambda\rho}\nabla_{\mu}\tau\nabla_{\nu}\tau\nabla^{\lambda}\tau\nabla^{\rho}\tau -G^{\mu}_{\nu}\nabla_{\mu}\tau\nabla^{\nu}\tau \Big]\Bigg),
\eea
where
\bea
G^{\mu\nu}_{\lambda\rho} &= & \epsilon^{\alpha_1\beta_1\alpha_2\beta_2\mu\nu}\epsilon_{\gamma_1\delta_1\alpha_2\beta_2\lambda\rho}R^{\gamma_1\delta_1}_{\alpha_1\beta_1}, \nonumber \\
G^{\mu}_{\nu} &= & \epsilon^{\alpha_1\beta_1\alpha_2\beta_2\gamma\mu}\epsilon_{\gamma_1\delta_1\gamma_2\delta_2\gamma\nu}R^{\gamma_1\delta_1}_{\alpha_1\beta_1}R^{\gamma_2\delta_2}_{\alpha_2\beta_2}. 
\eea
These interactions survive the flat space limit. After integrating by parts, the flat space limit gives,
\be \label{6danomalyaction}
S_{anomaly} = a\int d^6x \Big( -\frac{1}{8}(\partial \tau)^6 +\frac{1}{30}(\partial \tau)^2\Box(\partial \tau)^2 \Big).
\ee

\newpage

\begin{thebibliography}{10}

\bibitem{Witten:1995zh}
Edward Witten, {\em {Some comments on string dynamics}}, {\tt
  arXiv:hep-th/9507121} {\tt [hep-th]}.

\bibitem{Klebanov:1996un}
Igor~R. Klebanov and Arkady~A. Tseytlin, {\em {Entropy of near extremal black
  p-branes}}, Nucl.Phys. {\bf B475} (1996) 164--178, {\tt arXiv:hep-th/9604089}
  {\tt [hep-th]}.

\bibitem{Henningson:1998gx}
M.~Henningson and K.~Skenderis, {\em {The holographic Weyl anomaly}}, JHEP {\bf
  07} (1998) 023, {\tt arXiv:hep-th/9806087}.

\bibitem{Bastianelli:2000hi}
F.~Bastianelli, S.~Frolov, and Arkady~A. Tseytlin, {\em {Conformal anomaly of
  (2,0) tensor multiplet in six-dimensions and AdS / CFT correspondence}}, JHEP
  {\bf 0002} (2000) 013, {\tt arXiv:hep-th/0001041} {\tt [hep-th]}.

\bibitem{Harvey:1998bx}
Jeffrey~A. Harvey, Ruben Minasian, and Gregory~W. Moore, {\em {Non-abelian
  tensor-multiplet anomalies}}, JHEP {\bf 09} (1998) 004, {\tt
  arXiv:hep-th/9808060}.

\bibitem{Yi:2001bz}
Piljin Yi, {\em {Anomaly of (2,0) theories}}, Phys.Rev. {\bf D64} (2001)
  106006, {\tt arXiv:hep-th/0106165} {\tt [hep-th]}.

\bibitem{Kim:2011mv}
Hee-Cheol Kim, Seok Kim, Eunkyung Koh, Kimyeong Lee, and Sungjay Lee, {\em {On
  instantons as Kaluza-Klein modes of M5-branes}}, JHEP {\bf 1112} (2011) 031,
  54 pages, 2 figures. v2: references added, figure improved, added comments on
  self-dual string anomaly, added new materials on the symmetric phase index,
  other minor corrections, {\tt arXiv:1110.2175} {\tt [hep-th]}.

\bibitem{Bolognesi:2011rq}
Stefano Bolognesi and Kimyeong Lee, {\em {1/4 BPS String Junctions and $N^3$
  Problem in 6-dim (2,0) Superconformal Theories}}, Phys.Rev. {\bf D84} (2011)
  126018, 13 pages, 8 figures, jhep-style, references added, {\tt
  arXiv:1105.5073} {\tt [hep-th]}.

\bibitem{Bolognesi:2011nh}
\bysame, {\em {Instanton Partons in 5-dim SU(N) Gauge Theory}}, Phys.Rev. {\bf
  D84} (2011) 106001, {\tt arXiv:1106.3664} {\tt [hep-th]}.

\bibitem{Intriligator:2000eq}
Kenneth~A. Intriligator, {\em {Anomaly matching and a Hopf-Wess-Zumino term in
  6d, N=(2,0) field theories}}, Nucl.Phys. {\bf B581} (2000) 257--273, {\tt
  arXiv:hep-th/0001205} {\tt [hep-th]}.

\bibitem{Ganor:1998ve}
Ori Ganor and Lubos Motl, {\em {Equations of the (2,0) theory and knitted
  five-branes}}, JHEP {\bf 9805} (1998) 009, {\tt arXiv:hep-th/9803108} {\tt
  [hep-th]}.

\bibitem{Paban:1998ea}
Sonia Paban, Savdeep Sethi, and Mark Stern, {\em {Constraints from extended
  supersymmetry in quantum mechanics}}, Nucl.Phys. {\bf B534} (1998) 137--154,
  {\tt arXiv:hep-th/9805018} {\tt [hep-th]}.

\bibitem{Paban:1998qy}
\bysame, {\em {Supersymmetry and higher derivative terms in the effective
  action of Yang-Mills theories}}, JHEP {\bf 9806} (1998) 012, {\tt
  arXiv:hep-th/9806028} {\tt [hep-th]}.

\bibitem{Sethi:1999qv}
Savdeep Sethi and Mark Stern, {\em {Supersymmetry and the Yang-Mills effective
  action at finite N}}, JHEP {\bf 9906} (1999) 004, {\tt arXiv:hep-th/9903049}
  {\tt [hep-th]}.

\bibitem{Douglas:2010iu}
Michael~R. Douglas, {\em {On D=5 super Yang-Mills theory and (2,0) theory}},
  JHEP {\bf 1102} (2011) 011, {\tt arXiv:1012.2880} {\tt [hep-th]}.

\bibitem{Lambert:2010iw}
N.~Lambert, C.~Papageorgakis, and M.~Schmidt-Sommerfeld, {\em {M5-Branes,
  D4-Branes and Quantum 5D super-Yang-Mills}}, JHEP {\bf 1101} (2011) 083, {\tt
  arXiv:1012.2882} {\tt [hep-th]}.

\bibitem{Ho:2011ni}
Pei-Ming Ho, Kuo-Wei Huang, and Yutaka Matsuo, {\em {A Non-Abelian Self-Dual
  Gauge Theory in 5+1 Dimensions}}, JHEP {\bf 1107} (2011) 021, {\tt
  arXiv:1104.4040} {\tt [hep-th]}.

\bibitem{Singh:2011id}
Harvendra Singh, {\em {Super-Yang-Mills and M5-branes}}, JHEP {\bf 1108} (2011)
  136, 15 pages/ minor changes, to appear in JHEP, {\tt arXiv:1107.3408} {\tt
  [hep-th]}.

\bibitem{Samtleben:2011fj}
Henning Samtleben, Ergin Sezgin, and Robert Wimmer, {\em {(1,0) superconformal
  models in six dimensions}}, JHEP {\bf 1112} (2011) 062, {\tt arXiv:1108.4060}
  {\tt [hep-th]}.

\bibitem{Chu:2011fd}
Chong-Sun Chu, {\em {A Theory of Non-Abelian Tensor Gauge Field with
  Non-Abelian Gauge Symmetry G x G}}, {\tt arXiv:1108.5131} {\tt [hep-th]}.

\bibitem{Chu:2012um}
Chong-Sun Chu and Sheng-Lan Ko, {\em {Non-abelian Action for Multiple
  M5-Branes}}, {\tt arXiv:1203.4224} {\tt [hep-th]}.

\bibitem{Samtleben:2012mi}
Henning Samtleben, Ergin Sezgin, Robert Wimmer, and Linus Wulff, {\em {New
  superconformal models in six dimensions: Gauge group and representation
  structure}}, {\tt arXiv:1204.0542} {\tt [hep-th]}.

\bibitem{Czech:2011dk}
Bartlomiej Czech, Yu-tin Huang, and Moshe Rozali, {\em {Amplitudes for Multiple
  M5 Branes}}, {\tt arXiv:1110.2791} {\tt [hep-th]}.

\bibitem{Linander:2011jy}
Hampus Linander and Fredrik Ohlsson, {\em {(2,0) theory on circle fibrations}},
  * Temporary entry *, {\tt arXiv:1111.6045} {\tt [hep-th]}.

\bibitem{Gustavsson:2011ur}
Andreas Gustavsson, {\em {A preliminary test of Abelian D4-M5 duality}},
  Phys.Lett. {\bf B706} (2011) 225--227, {\tt arXiv:1111.6339} {\tt [hep-th]}.

\bibitem{Young:2011aa}
Donovan Young, {\em {Wilson Loops in Five-Dimensional Super-Yang-Mills}}, JHEP
  {\bf 1202} (2012) 052, 1+18 pages, 3 figures. v2 added a reference and made
  minor cosmetic changes, JHEP version. v3 added generalization to arbitrary
  smooth, closed contours, added references. v4 added references and clarified
  discussion beneath eq. (23), {\tt arXiv:1112.3309} {\tt [hep-th]}.

\bibitem{Berman:2008be}
David~S. Berman, Laura~C. Tadrowski, and Daniel~C. Thompson, {\em {Aspects of
  Multiple Membranes}}, Nucl. Phys. {\bf B802} (2008) 106--120, {\tt
  arXiv:0803.3611} {\tt [hep-th]}.

\bibitem{Berman:2009xd}
David~S. Berman, Malcolm~J. Perry, Ergin Sezgin, and Daniel~C. Thompson, {\em
  {Boundary Conditions for Interacting Membranes}}, JHEP {\bf 1004} (2010) 025,
  {\tt arXiv:0912.3504} {\tt [hep-th]}.

\bibitem{Berman:2009kj}
David~S. Berman and Daniel~C Thompson, {\em {Membranes with a boundary}}, Nucl.
  Phys. {\bf B820} (2009) 503--533, {\tt arXiv:0904.0241} {\tt [hep-th]}.

\bibitem{Terashima:2010ji}
Seiji Terashima and Futoshi Yagi, {\em {On Effective Action of Multiple
  M5-branes and ABJM Action}}, JHEP {\bf 1103} (2011) 036, {\tt
  arXiv:1012.3961} {\tt [hep-th]}.

\bibitem{Lambert:2011eg}
N.~Lambert, H.~Nastase, and C.~Papageorgakis, {\em {5D Yang-Mills instantons
  from ABJM Monopoles}}, Phys.Rev. {\bf D85} (2012) 066002, 29 pages, Latex/
  v2: added references and a comment on the graviphoton coupling in section 5/
  v3: typos corrected and references added, {\tt arXiv:1111.5619} {\tt
  [hep-th]}.

\bibitem{Papageorgakis:2011xg}
Constantinos Papageorgakis and Christian Saemann, {\em {The 3-Lie Algebra (2,0)
  Tensor Multiplet and Equations of Motion on Loop Space}}, JHEP {\bf 1105}
  (2011) 099, {\tt arXiv:1103.6192} {\tt [hep-th]}.

\bibitem{Berman:2007bv}
David~S. Berman, {\em {M-theory branes and their interactions}}, Phys. Rept.
  {\bf 456} (2008) 89--126, {\tt arXiv:0710.1707} {\tt [hep-th]}.

\bibitem{Keurentjes:2002dc}
Arjan Keurentjes and Savdeep Sethi, {\em {Twisting E8 five-branes}}, Phys.Rev.
  {\bf D66} (2002) 046001, {\tt arXiv:hep-th/0205162} {\tt [hep-th]}.

\bibitem{Tachikawa:2011ch}
Yuji Tachikawa, {\em {On S-duality of 5d super Yang-Mills on $S^1$}}, JHEP {\bf
  1111} (2011) 123, {\tt arXiv:1110.0531} {\tt [hep-th]}.

\bibitem{Paban:1998mp}
Sonia Paban, Savdeep Sethi, and Mark Stern, {\em {Summing up instantons in
  three-dimensional Yang-Mills theories}}, Adv.Theor.Math.Phys. {\bf 3} (1999)
  343--361, Published in Adv.Theor.Math.Phys. 3 issue 2, {\tt
  arXiv:hep-th/9808119} {\tt [hep-th]}.


\bibitem{LIE}
Marc A.~A. van Leeuwen, Arjeh Cohen, and Bert Lisser, {\em {http://www-math.univ-poitiers.fr/$\sim$maavl/LiE/factsheet.html}}.


\bibitem{Hyun:1999hf}
Seungjoon Hyun, Youngjai Kiem, and Hyeonjoon Shin, {\em {Supersymmetric
  completion of supersymmetric quantum mechanics}}, Nucl.Phys. {\bf B558}
  (1999) 349, {\tt arXiv:hep-th/9903022} {\tt [hep-th]}.

\bibitem{Kazama:2002jm}
Y.~Kazama and T.~Muramatsu, {\em {Power of supersymmetry in D particle
  dynamics}}, Nucl.Phys. {\bf B656} (2003) 93--131, 44 pages, v2: typos
  corrected, published version Report-no: UT-Komaba 02-11, {\tt
  arXiv:hep-th/0210133} {\tt [hep-th]}.

\bibitem{Sethi:2004np}
Savdeep Sethi, {\em {Structure in supersymmetric Yang-Mills theory}}, JHEP {\bf
  0410} (2004) 001, {\tt arXiv:hep-th/0404056} {\tt [hep-th]}.

\bibitem{Buchbinder:2001xy}
I.L. Buchbinder and E.A. Ivanov, {\em {Complete N=4 structure of low-energy
  effective action in N=4 superYang-Mills theories}}, Phys.Lett. {\bf B524}
  (2002) 208--216, {\tt arXiv:hep-th/0111062} {\tt [hep-th]}.

\bibitem{Belyaev:2011wa}
Dmitry~V. Belyaev and Igor~B. Samsonov, {\em {Bi-harmonic superspace for N=4
  d=4 super Yang-Mills}}, JHEP {\bf 09} (2011) 056, {\tt arXiv:1106.0611} {\tt
  [hep-th]}.

\bibitem{Belyaev:2011dg}
\bysame, {\em {Wess-Zumino term in the N=4 SYM effective action revisited}},
  JHEP {\bf 04} (2011) 112, {\tt arXiv:1103.5070} {\tt [hep-th]}.

\bibitem{Buchbinder:2011zu}
I.~L. Buchbinder, E.~A. Ivanov, I.~B. Samsonov, and B.~M. Zupnik, {\em
  {Superconformal N=3 SYM Low-Energy Effective Action}}, JHEP {\bf 01} (2012)
  001, {\tt arXiv:1111.4145} {\tt [hep-th]}.

\bibitem{Dine:1997nq}
Michael Dine and Nathan Seiberg, {\em {Comments on higher derivative operators
  in some SUSY field theories}}, Phys.Lett. {\bf B409} (1997) 239--244, {\tt
  arXiv:hep-th/9705057} {\tt [hep-th]}.

\bibitem{Buchbinder:1998qd}
E.I. Buchbinder, I.L. Buchbinder, and S.M. Kuzenko, {\em {Nonholomorphic
  effective potential in N=4 SU(n) SYM}}, Phys.Lett. {\bf B446} (1999)
  216--223, {\tt arXiv:hep-th/9810239} {\tt [hep-th]}.

\bibitem{Dine:1999jq}
Michael Dine and Josh Gray, {\em {Nonrenormalization theorems for operators
  with arbitrary numbers of derivatives in N=4 Yang-Mills theory}}, Phys.Lett.
  {\bf B481} (2000) 427--435, Preliminary version, {\tt arXiv:hep-th/9909020}
  {\tt [hep-th]}.

\bibitem{Nakayama:2011wq}
Yu~Nakayama, {\em {On $\epsilon$-conjecture in $a$-theorem}}, {\tt
  arXiv:1110.2586} {\tt [hep-th]}.

\bibitem{Schwimmer:2010za}
A.~Schwimmer and S.~Theisen, {\em {Spontaneous Breaking of Conformal Invariance
  and Trace Anomaly Matching}}, Nucl. Phys. {\bf B847} (2011) 590--611, {\tt
  arXiv:1011.0696} {\tt [hep-th]}.

\bibitem{Komargodski:2011vj}
Zohar Komargodski and Adam Schwimmer, {\em {On Renormalization Group Flows in
  Four Dimensions}}, JHEP {\bf 12} (2011) 099, {\tt arXiv:1107.3987} {\tt
  [hep-th]}.

\bibitem{Oliva:2010zd}
Julio Oliva and Sourya Ray, {\em {Classification of Six Derivative Lagrangians
  of Gravity and Static Spherically Symmetric Solutions}}, Phys. Rev. {\bf D82}
  (2010) 124030, {\tt arXiv:1004.0737} {\tt [gr-qc]}.

\bibitem{Deser:1993yx}
Stanley Deser and A.~Schwimmer, {\em {Geometric classification of conformal
  anomalies in arbitrary dimensions}}, Phys. Lett. {\bf B309} (1993) 279--284,
  {\tt arXiv:hep-th/9302047}.

\bibitem{Arakelian:1995ye}
T.~Arakelian, D.~R. Karakhanian, R.~P. Manvelyan, and R.~L. Mkrtchian, {\em
  {Trace anomalies and cocycles of the Weyl group}}, Phys. Lett. {\bf B353}
  (1995) 52--56.

\end{thebibliography}

\ifx\undefined\bysame
\newcommand{\bysame}{\leavevmode\hbox to3em{\hrulefill}\,}
\fi

\end{document}